\PassOptionsToPackage{table}{xcolor}
\documentclass[11pt]{article} 

\usepackage{authblk}

\usepackage{changepage}
\usepackage[justification=centering]{caption}
\usepackage{url}
\usepackage[margin=0.8in]{geometry}
\usepackage[T1]{fontenc}
\usepackage{tgtermes}  

\usepackage{amsmath}
\usepackage{newtxmath}  
\usepackage{setspace}
\usepackage[tiny,compact]{titlesec}

\usepackage{placeins}

\usepackage[numbers]{natbib}
\usepackage[final]{microtype}
\usepackage{xcolor}
\usepackage{float}

\usepackage{textgreek}

\newcommand{\comment}[1]{}

\usepackage{graphicx}
\graphicspath{ {./images} }
\usepackage[utf8]{inputenc}
\usepackage[english]{babel}

\usepackage{tikz}
\usetikzlibrary{shapes.geometric, arrows, fit, positioning}

\tikzstyle{mainb} = [rectangle, rounded corners, minimum width=3cm, minimum height=1cm, text centered, draw=black, fill=blue!20, text width=3cm]
\tikzstyle{arrow} = [thick,->,>=stealth]

\usepackage{makecell}

\usepackage{pdflscape}
\usepackage{longtable}
\usepackage{multirow}
\usepackage{enumitem}

\usepackage{array}

\usepackage{hyperref}
\usepackage{tabularx}
\usepackage{adjustbox}
\usepackage{hhline}
\usepackage{tcolorbox}

\definecolor{headercolor}{rgb}{0.772, 0.796, 0.847}
\definecolor{altrowcolor}{rgb}{0.882, 0.894, 0.922}
\definecolor{lighteraltrowcolor}{rgb}{0.953, 0.957, 0.965}

\title{The Anatomy of Coronary Risk: How Arterial Geometry Shapes Coronary Artery Disease through Blood Flow Haemodynamics – Latest Methods, Insights and Clinical Implications}

\author{
C. Shen \textsuperscript{1,*}, M. Zhang \textsuperscript{1}, H. Keramati \textsuperscript{1}, S. Zhang \textsuperscript{1}, R. Gharleghi \textsuperscript{1}, J. J. Wentzel \textsuperscript{2}, M. O. Khan \textsuperscript{3}, U. Morbiducci \textsuperscript{4}, A. Qayyum \textsuperscript{5}, S. A. Niederer \textsuperscript{5}, S. Samant \textsuperscript{6}, Y. S. Chatzizisis \textsuperscript{7}, D. Almeida \textsuperscript{8}, Tsung-Ying Tsai \textsuperscript{9}, P. Serruys \textsuperscript{9}, S. Beier \textsuperscript{1}\\
\textsuperscript{1} School of Mechanical and Manufacturing Engineering, University of New South Wales, Sydney NSW 2052, Australia\\
\textsuperscript{2} Biomedical Engineering, Cardiology Department, ErasmusMC, Rotterdam, The Netherlands\\
\textsuperscript{3} Department of Electrical, Computer, and Biomedical Engineering, Toronto Metropolitan University, Toronto, Canada\\
\textsuperscript{4} Department of Mechanical and Aerospace Engineering, Politecnico di Torino, Turin, Italy\\
\textsuperscript{5} Imperial College London\\
\textsuperscript{6} Department of Medicine, Montefiore Medical Center, Bronx, New York\\
\textsuperscript{7} Centre for Digital Cardiovascular Innovations, Division of Cardiovascular Medicine, Miller School of Medicine, University of Miami, Miami, Florida\\

\textsuperscript{8} Virtonomy GmbH, Paul-Heyse-Straße 6, 80336 Munich, Germany\\
\textsuperscript{9} Corrib research centre for advanced imaging and core laboratory, University of Galway\\

*Corresponding author\\
Email: chi.shen@student.unsw.edu.au
}
\date{}

\begin{document}

\maketitle

\begin{tcolorbox}[sharp corners, rounded corners=southeast, colback=altrowcolor,colframe=altrowcolor]

\textbf{Key points}
\begin{itemize}
\item Despite being the leading cause of global morbidity and mortality, the assessment of individual risk throughout the course of Coronary Artery Disease (CAD) remains sub-optimal.
\item Coronary vascular anatomy, with high variation among individuals, drives local haemodynamics that affects endothelial cellular health underlying CAD, suggesting that coronary anatomy may be a useful surrogate marker to refine and personalise CAD risk assessments at all stages of the disease.
\item The field increasingly focuses on haemodynamic analysis based on patient-specific coronary anatomy, considering increasing ability to characterise anatomical details and larger-scale analysis. 
\item The understanding of the combined effect of anatomical features on CAD remains limited. The main reasons for the ambiguity are the challenges of \textit{in vivo} validation, the need for more whole coronary artery tree analyses,  a shortage of large-scale, population-specific studies, and the need for more precise and consistent anatomical feature definitions.
\item Several opportunities exist: (1) Enhance patient-specific coronary datasets by incorporating broader, population-specific factors (including boundary conditions) using advanced imaging technologies. (2)  Develop innovative machine learning approaches to automate coronary artery segmentation and haemodynamic analysis on a larger scale. (3) Exploration of the interdependent effects of all anatomical features for the whole artery trees on the local haemodynamics under different conditions. 
\end{itemize}
\end{tcolorbox}

\newpage
\section*{Abstract}

Despite tremendous advances in cardiovascular medicine, significant opportunities remain to improve coronary artery disease (CAD) prevention and treatment strategies. The key limitation lies in the understanding of disease formation and progression mechanisms. The coronary anatomy plays an important role in local haemodynamics, governing endothelial health and, thus, pathophysiological responses. The significant variation of the coronary anatomy among patients, with significant trends across different populations, increases the complexity of understanding the details of disease progression. This review covers different aspects of the current status and understanding of the blood flow investigation in coronary arteries. We summarised the current knowledge of the haemodynamic effect of coronary anatomy and its evaluation and analysis methods. We discussed recent progress across medical imaging techniques and computational haemodynamic analysis. Based on the reviewed papers, we identified the persisting knowledge gaps and challenges in the field. We then elaborated on future directions and opportunities to increase understanding of the fundamental mechanism of CAD in individuals representative of large populations and how this may translate to the patient's bedside.  

\section{Introduction}

Coronary Artery Disease (CAD) continues to increase the social and economic burden on a global scale as a leading cause of morbidity and mortality marked by significant socio-economic disparities \cite{Association2023-CADstatistics, Vaduganathan2022}. Atherosclerotic plaque is the most common type of CAD, characterised by a chronic and progressive pathology \cite{agrawal2020CAD}, whereby the coronary blood vessels are narrowed through atherosclerosis development,  progressively leading to the insufficient blood supply to the heart muscle, causing shortness of breath, chest pain, heart failure, and even death \cite{Bhatia2010CoronaryArteryDisease}. 

Interestingly, atherosclerotic disease is more likely to form and progress in coronary segments with complex anatomy \cite{cunningham2005}, as blood flow-induced shear forces at the vessel wall affect the endothelial function \cite{tamargo2023flow_endothelial}. In fact, the concept of the `Anatomy of Risk' was initially proposed by Friedman et al. \cite{Friedman1983anatomyrisk} in 1983, suggesting that the vascular geometrical features might contribute to arterial diseases among individuals and populations. The regions around bifurcations, and segments with severe curvature are at a high risk of pathophysiological processes \cite{cunningham2005, Temov2016bifurcation_CAD,Tuncay2018invivo_tortuosity_CTA, Eleid2014tortuosity}. Computational studies have confirmed the distributions of adverse haemodynamics in these areas \cite{BeierBifurcation2016,Chiastra2017Healthy_and_diseased,Shen2021Secondary_flow} and demonstrated that the effects of anatomical characteristics, such as bifurcation angle, vessel curvature, and diameter, are interdependent, exerting a combined influence on local flow patterns. \cite{BeierBifurcation2016}. Understanding the anatomy of risk can improve current risk assessment strategies by identifying individuals without Standard Modifiable Risk Factors (SMuRFs) \cite{Esau2022risk_evaluation,Kong2023}. Patients without SMuRFs account for an increasing proportion of acute coronary syndrome cases and exhibit higher in-hospital mortality and cardiogenic shock rates. \cite{Kong2023}. Despite the recent advancements, the anatomy of risk has yet to be fully understood, thus limiting its clinical applications to date. 

Moreover, blood flow characteristics are intricately affected by the local coronary anatomy. Anatomical features are among the determining factors in disease onset and progression through their influence on haemodynamics \cite{cunningham2005}. In recent years, increasing efforts have been made to identify patient-specific coronary arterial vessel characteristics that result in adverse haemodynamics. Additionally, understanding and classifying arterial anatomical remodelling with age and disease stage would further elucidate this dynamic that has proven challenging since coronary arteries vary widely between individuals \cite{Pau2017shape_in_normal_population} and between population groups \cite{Skowronski2020population_difference}. 

In recent years, more complex, large-scale considerations beyond a few patient-specific cases are now in the realms of possibility due to increased access and lower cost of computational resources, more advanced imaging techniques \cite{mezquita2023clinical_image}, an increasing number of open-access datasets \cite{Gharleghi2023ASOCADATA, li2025medshapenet,Zeng2023_opendata}. Furthermore, improved (although not yet standardised) segmentation capacities \cite{Gharleghi2022_segmentation_review} have increased accessibility and streamlined subsequent in-depth computational blood flow analysis efforts. By integrating computational methods with clinical information, the understanding of CAD mechanisms continues to advance. This strategy is attracting growing interest from cardiovascular clinicians and shows strong promise for future clinical application \cite{Chiastra2023CFDsurvey}.

\begin{figure}
    \centering
    \includegraphics[width=1\linewidth]{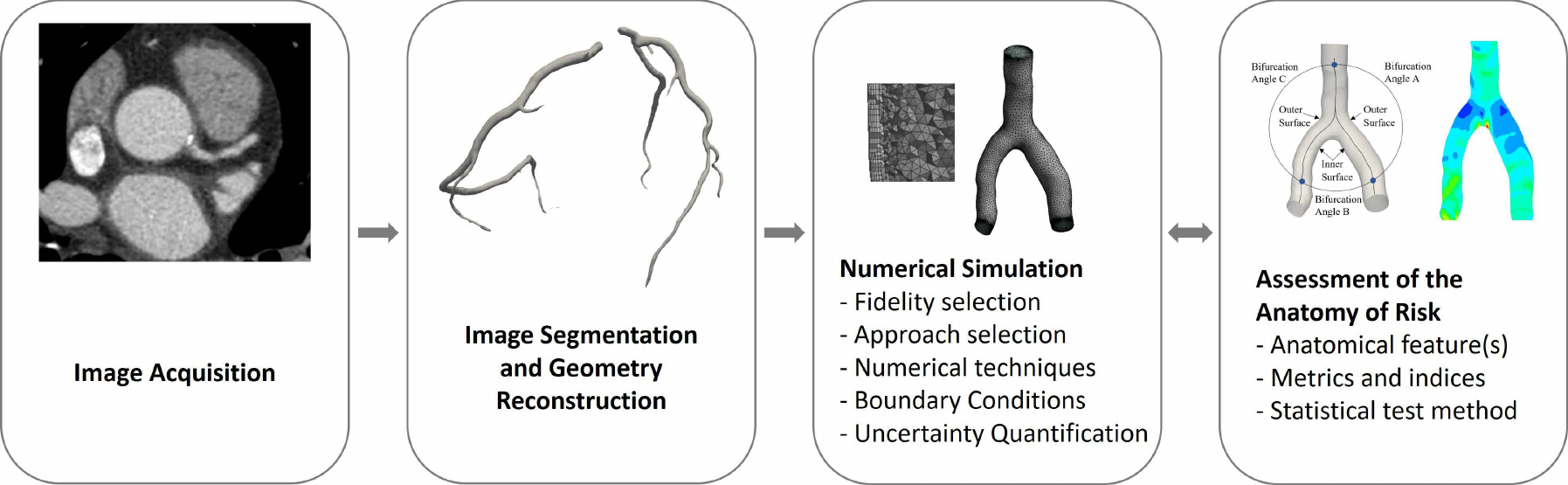}
    \caption{The main steps to investigate the anatomy of risk. The two-sided arrow between the two last blocks indicates the interaction between the numerical approaches selection and the assessment of the anatomical feature(s) effects.}
    \label{fig: steps}
\end{figure}

The assessment of the effect of the anatomical and haemodynamic characteristics on CAD comprises several steps (Figure \ref{fig: steps}), each outlined here. First, we provide an overview of the rapidly developing field of the latest invasive and non-invasive imaging techniques suitable for coronary tree imaging, followed by the current and emerging segmentation methods. Next, we discuss different numerical approaches for analysing CAD, focusing on current best practices and the latest developments and opportunities. We then discuss the `Anatomy of Risk' hypothesis \cite{Friedman1983anatomyrisk}, its reappraisal in the context of these latest technological developments and approaches, their latest findings and how this may translate into clinical practice today and in future by highlighting the remaining challenges and resulting opportunities. 

\FloatBarrier
\section{Image Acquisition and Reconstruction}
Relevant imaging acquisition plays a crucial role in coronary artery research. This section addresses the latest developments in imaging techniques and image post-processing relevant to this field.

\subsection{Coronary Artery Imaging Technologies}

A range of coronary imaging techniques to delineate coronary anatomy are available (Table \ref{table_image}), which can be categorised into non-invasive (Computed Tomographic Coronary Angiography; CTCA and Magnetic Resonance Angiography; MRA), and invasive techniques (Invasive Coronary Angiography; ICA, Optical Coherence Tomography; OCT, Intravascular Ultrasound; IVUS). Invasive imaging is generally more costly and has a higher patient risk associated with it \cite{DISCHARGE2022_CT_or_ICA}, although it traditionally provides higher resolution, thus, better diagnostic insights than non-invasive imaging. Recently, the use of non-invasive imaging modalities is increasingly recognised in its capacity to evaluate the extent and composition of plaques in coronary arteries whilst being a more conservative and less expensive approach than invasive methods \cite{Serruys2021CTCA}. Moreover, non-invasive techniques enable new assessment features to be incorporated, directly relevant for risk stratification of CAD, such as epicardial adipose tissue linked to plaque development and arrhythmias \cite{Patel2022fibrillation,Conte2022ArrhythmiasandFibrillation}, and myocardial fibrosis assessed through Extracellular Volume (ECV) \cite{Cundari2023_ECV}. These new features can also realise a more in-depth plaque characterisation for assessment of plaque vulnerability with the latest ultra-high spatial resolution techniques \cite{Schuijf2022image_high_resolution}. Recently, AI-augmented imaging software, such as Vitrea \cite{Vitrea}, Tera Reco \cite{terarecon}, Artrya \cite{artrya}, Cleerly \cite{cleerly2024} and Medis \cite{medis}, has emerged, providing machine learning-trained vessel wall assessments and enhancing risk stratification capabilities. 

However, direct haemodynamic assessment is still limited, as any minor luminal boundary uncertainties can significantly affect the efficacy of the resulting analyses \cite{Kweon2018lumen_reconstruction}. Recent clinical studies have attempted to push these current boundaries, whereby the EMERALD trial \cite{Lee2019HighRiskPlaquesACS}, combining Computational Fluid Dynamics (CFD) with CTCA to predict coronary plaque rupture, developed a risk predictor superior in identifying vulnerable plaques when compared to the conventional CTCA-derived plaque dimension and characteristics assessment for both primary and secondary prevention \cite{Lee2019HighRiskPlaquesACS}. Thus, the modality of choice to date remains non-invasive Computed Tomography (CT) considering all trade-offs, providing vascular anatomy and limited plaque calcification information \cite{Eckert2015imaging_review_ICA} and is far superior in resolution compared to the only other non-invasive coronary imaging option, Magnetic Resonance Imaging (MRI). 

Nevertheless, CTCA requires extensive post-processing and exposes patients to radiation \cite{Lin2019non_invasive_image,Tarkin2016Imaging_review}. MRI techniques, even with current spatial and temporal limitations, can provide artery structure, plaque diagnosis,microvascular dysfunction, myocardium perfusion, and information about blood velocity and volume flow rate without the need for contrast injection or radiation \cite{Tarkin2016Imaging_review}. It is particularly useful for patients with contraindications to iodinated contrast or those requiring repeated imaging over time \cite{Tarkin2016Imaging_review}. Overall, in the long-term pursuit of a comprehensive imaging evaluation that has the potential to serve as a single non-invasive, safe, and convenient approach for diagnostics, risk management, and treatment decisions, the question remains whether CT-based methods will prevail or MRI techniques will outperform them. CT-based methods require radiation, while MRI techniques face persisting challenges of providing higher temporal and spatial resolution, motion artefact with longer scan times, being still highly costly, and having limited accessibility since it precludes many patients due to claustrophobia or metal implants \cite{Owen2011Imaging_review,Kato2020MRA,Androulakis2022MRA}. A Compressed Sensing Artificial Intelligence (CSAI) framework has been developed to overcome the limitations of MRA by accelerating acquisition time and enhancing image quality \cite{DeepLearning_MRA2023}. 

ICA remains the gold standard for coronary stenosis detection. It has been more accurate than its non-invasive counterpart, CTCA \cite{Mantella2021image_review}, particularly as the latest ultrahigh-resolution scanners are still not widely available. However, ICA does not provide information about plaque composition or vessel wall characteristics, which limits its role in assessing plaque vulnerability. 

For high-risk patients requiring detailed coronary assessment beyond luminal narrowing, intravascular imaging techniques such as IVUS and OCT provide superior insights. OCT has a ten times greater resolution than IVUS and is thus ideal for analysing vulnerable plaque and assessing post-stenting complications, including stent malapposition and neointimal hyperplasia \cite{Barbieri2022OCT}. IVUS has greater vessel penetration up to 2.7 times than OCT, making it more suitable for analysing deep plaque burden, vessel remodelling, and overall plaque burden quantification \cite{Barbieri2022OCT,Nagaraja2020IVUS_OCT}.  Spectral Virtual Histology (VH-IVUS) analysis \cite{Nair2002IVUS} is the gold standard for plaque quantification and characterisation, enabling plaque border and volume detection and assessment of vessel remodelling - both positive (outward) and negative (inward the lumen) - across intima, media, and adventitia \cite{Roleder2015IVUS,Schoenhagen2001remodeling,Nissen2001IVUS}. 

Advanced imaging technologies have been developed to address the limitations. For example, the radiation exposure associated with CTCA has been significantly reduced by recent advancements such as iterative reconstruction techniques and photon-counting CT, which lower radiation dose while improving image quality, enabling higher imaging resolution and less blooming artefact \cite{photon_counting_CTCAMeloni2024}. Additionally, hybrid imaging systems integrate multiple modalities to leverage their complementary advantages. The integrated IVUS-OCT catheter combines IVUS and OCT for enhanced vascular imaging \cite{Ono2020IVUS_OCT}. NIRS-IVUS integrates NIRS (Near-Infrared Spectroscopy) to enhance the differentiation of plaque rupture, plaque erosion, and calcified nodules while maintaining deep tissue penetration \cite{NIRS_IVUS_Terada2021}. The novel DeepOCT-NIRS multimodal system enables high-resolution intracoronary imaging with improved depth penetration and co-registered NIRS-based lipid identification \cite{Ali2024_deepOCT_IVUS}. Positron Emission Tomography (PET)/CTCA enables a non-invasive assessment of plaque characteristics, including plaque inflammation and vulnerability \cite{pet_CTCA_Majeed2021,Kwiecinski2023}. However, these technologies are not yet widely adopted in clinical practice and remain primarily confined to research settings. For patient diagnostics, imaging techniques in Table \ref{table_image} are still the preferred approach due to factors such as accessibility, cost, and ease of use. It is important to note that imaging techniques are on the verge of being optimised through AI enhancements, which will allow significant strides in medical imaging, diagnosis, and thus prevention in the near future. Specifically, AI-driven plaque characterisation, automated stenosis quantification, and machine learning-based haemodynamic risk stratification are emerging applications that are expected to significantly improve diagnostic accuracy and accelerate clinical decision-making \cite{Follmer2024_AI_vulnerable_plaque,nagayama2023improving,Su2020ML_CFD, Elias2024_AI_CV_Part1, Jain2024_AI_CV_Part2}.

\begin{landscape}
\begin{longtable}{|l|l|m{4cm}|m{4cm}|m{4cm}|m{4cm}|}
\caption{Advantages, limitations and optimal clinical applications of coronary imaging techniques based on current evidence and clinical trends.} \label{table_image} \\

\hline
\rowcolor{headercolor}
\textbf{Types} & \textbf{Techniques} & \textbf{Advantages} & \textbf{Limitations} & \textbf{Clinical Scenario} & \textbf{AI-assisted tools} \\
\hline
\endfirsthead

\hline
\rowcolor{headercolor}
\textbf{Types} & \textbf{Techniques} & \textbf{Advantages} & \textbf{Limitations} & \textbf{Clinical Scenario} & \textbf{AI-assisted tools} \\
\hline
\endhead

\hline
\endfoot

\hline
\endlastfoot

\rowcolor{lighteraltrowcolor}
\multirow{2}{*}{\textbf{Non-invasive}} & 
\makecell[{{p{3.5cm}}}]{\includegraphics[width=3.2cm]{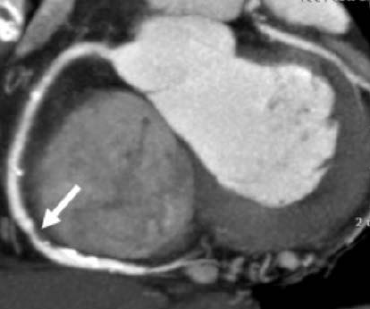}\\Computed Tomography Coronary Angiogram \\(CTCA)\cite{arbab2011ICA_VS_CTCA}} &  
\begin{itemize}[leftmargin=*,itemsep=0pt,parsep=0pt]
    \item High resolution (0.5 mm)
    \item Short scanning time (~10 min)
    \item Semi-quantitative plaque characteristics assessment
    \item 3D coronary anatomy
\end{itemize} & 
\begin{itemize}[leftmargin=*,itemsep=0pt,parsep=0pt]
    \item High radiation exposure
\end{itemize} &
\begin{itemize}[leftmargin=*,itemsep=0pt,parsep=0pt]
    \item First-line for stable CAD diagnosis
    \item Limited plaque assessment
\end{itemize} &
Machine learning models for automated plaque characterisation (e.g., Vitrea, Artrya); AI-driven FFR-CT for non-invasive haemodynamic assessment (HeartFlow, Inc.) \\

\hhline{~-|-|-|-|-}

\rowcolor{lighteraltrowcolor}
 & 
\makecell[{{p{3.5cm}}}]{\includegraphics[width=3.2cm]{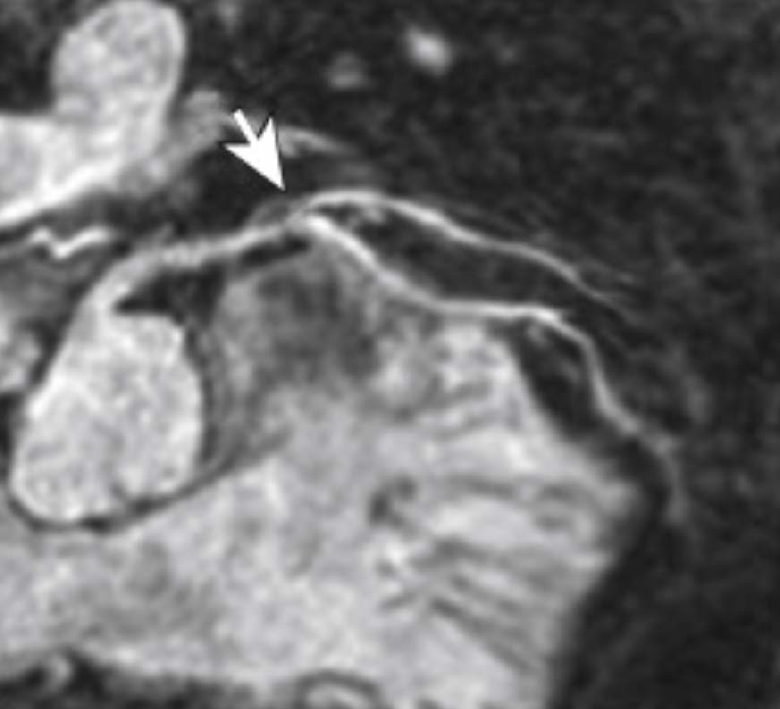}  \\ Magnetic Resonance Angiography \\(MRA)\cite{Henningsson2017_MRA}}&  
\begin{itemize}[leftmargin=*,itemsep=0pt,parsep=0pt]
    \item Radiation-free
    \item Functional assessment of myocardium
    \item 3D coronary anatomy
\end{itemize} & 
\begin{itemize}[leftmargin=*,itemsep=0pt,parsep=0pt]
    \item Low resolution (~1.0 mm)
    \item Long scanning time (~30 min)
    \item Motion artefact
\end{itemize} &
\begin{itemize}[leftmargin=*,itemsep=0pt,parsep=0pt]
    \item Patients with renal dysfunction
    \item Repeated imaging needs
\end{itemize} &
Compressed Sensing AI (CSAI)-based tools (Siemens Healthineers) and AIR™ Recon DL (GE Healthcare) are emerging AI models for plaque and perfusion analysis \\

\hline
\rowcolor{altrowcolor}
\multirow{3}{*}{\textbf{Invasive}} & 
\makecell[{{p{3.5cm}}}]{\includegraphics[width=3.2cm]{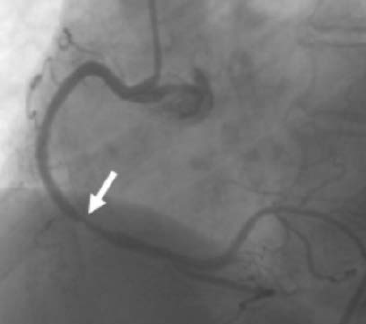} \\Invasive Coronary Angiography \\(ICA)\cite{arbab2011ICA_VS_CTCA}} &  
\begin{itemize}[leftmargin=*,itemsep=0pt,parsep=0pt]
    \item High resolution (~0.2 mm)
    \item Precise stenosis severity assessment (Gold standard)
\end{itemize} & 
\begin{itemize}[leftmargin=*,itemsep=0pt,parsep=0pt]
    \item High radiation exposure
    \item Inability to characterise plaque composition
    \item 2D projection images only
\end{itemize} &
\begin{itemize}[leftmargin=*,itemsep=0pt,parsep=0pt]
    \item Gold standard for CAD diagnosis
    \item Acute coronary syndrome
    \item Urgent PCI planning
\end{itemize} &
AI-assisted QCA (Quantitative Coronary Angiography) tools for automated lesion severity and vessel sizing (e.g., Medis QAngio XA, Pie Medical CAAS QCA, CathWorks for physiology-assisted reconstruction) \\

\hhline{~-|-|-|-|-}

\rowcolor{altrowcolor}
 & 
\makecell[{{p{3.5cm}}}]{\includegraphics[width=3.2cm]{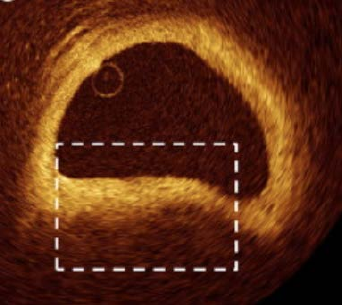} \\Optical Coherence Tomography \\(OCT)\cite{Bezerra2009OCT}} &  
\begin{itemize}[leftmargin=*,itemsep=0pt,parsep=0pt]
    \item Ultra-high resolution (Axial: ~15 \(\mu\)m; Lateral: ~50 \(\mu\)m)
    \item Plaque structure assessment
\end{itemize} & 
\begin{itemize}[leftmargin=*,itemsep=0pt,parsep=0pt]
    \item Low penetration depth (< 3 mm)
    \item 2D cross-sectional images only
\end{itemize} &
\begin{itemize}[leftmargin=*,itemsep=0pt,parsep=0pt]
    \item Vulnerable plaque detection
    \item Stent evaluation (malapposition, neointima)
\end{itemize} &
AI algorithms for automated stent apposition and neointimal coverage assessment (e.g., Ultreon™ 1.0 Software by Abbott Laboratories; emerging CNN-based tools for plaque morphology analysis) \\

\hhline{~-|-|-|-|-}

\rowcolor{altrowcolor}
 & 
\makecell[{{p{3.5cm}}}]{\includegraphics[width=3.2cm]{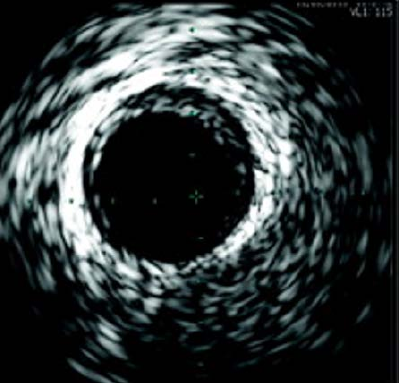} \\Intravascular Ultrasound \\(IVUS)\cite{Wang2023IVUS}} &  
\begin{itemize}[leftmargin=*,itemsep=0pt,parsep=0pt]
    \item Very-high resolution (Axial: ~150 \(\mu\)m; Lateral: ~200 \(\mu\)m)
    \item High penetration depth (> 8 mm)
    \item Plaque characteristics assessment (Gold standard)
\end{itemize} & 
\begin{itemize}[leftmargin=*,itemsep=0pt,parsep=0pt]
    \item 2D cross-sectional images only
\end{itemize} &
\begin{itemize}[leftmargin=*,itemsep=0pt,parsep=0pt]
    \item Deeper plaque burden evaluation
    \item Vessel remodelling analysis
\end{itemize} &
AI-assisted tools for plaque composition and vessel segmentation (e.g., Medis QIvus, CAAS IVUS by Pie Medical; deep learning-based segmentation models in research) \\

\end{longtable}
\end{landscape}

\FloatBarrier

\subsection{Coronary Geometry Segmentation}

Coronary images require post-processing to reconstruct and delineate the three-dimensional luminal structures.. For cardiac volume scanning techniques (e.g. CTCA or MRA), standard post-processing protocols, such as curved Multi-Planar Reconstruction (MPR) and Volume Rendering \cite{Rampidis2022} enable semi-quantitative characterisation of the stenoses and plaques, whereby the 3D computational models of the coronary tree can be reconstructed by various manual or automated methods \cite{Gharleghi2022_segmentation_review}. During intravascular imaging (e.g. IVUS or OCT), 2D arterial cross-sections are acquired. Fusion techniques combine multiple imaging techniques to enable a simultaneous assessment of the overall coronary anatomy with high-resolution plaque and vessel wall characteristics, for example, ICA combined with OCT \cite{Poon2023ICA_OCT}, or CTCA combined with IVUS \cite{Kilic2020reconstruction_review,DeNisco2024PlaqueProgression}, through calibration and registration along the vascular centreline.

Moreover, segmentation and 3D reconstruction of the coronary tree from CTCA images \cite{Gharleghi2022_segmentation_review}, and other methods such as ICA \cite{He2025_ICA_segmentation}, OCT \cite{Balaji2021, Lee2020OCT_segmentation} and IVUS \cite{Bajaj2021IVUS}, have markedly progressed over the last decade. Previously, the key segmentation processes relied heavily on manual verification, utilising semi-automated segmentation tools such as open-source libraries, e.g. 3D Slicer \cite{Fedorov2012Slicer}, ImageJ \cite{Rueden2017}, Blender \cite{blender2023}, Vascular Modelling Toolkit (VMTK) \cite{Antiga2008_VMTK}, Simvascular \cite{simvascular} or commercial software, e.g. Materialize  Mimics \cite{Mimics} or OsiriX \cite{Rosset2004}. AI-driven methods, particularly deep learning-based segmentation, have increasingly complemented conventional approaches.

This labour-intensive and time-consuming procedure has recently transitioned to more automated approaches, primarily based on fully supervised deep learning methods that utilise annotated datasets for training. Full automated voxel-wise neural networks have increasingly been used for coronary artery segmentation since 2017 \cite{Gharleghi2022_segmentation_review}. However, fully supervised deep learning requires large, annotated datasets, which can be costly and time-consuming to obtain. To address this limitation, semi-supervised learning has emerged, leveraging a small amount of labelled data alongside a larger pool of unlabelled data to improve model generalisation. Multi-modality learning integrates information from different imaging techniques (e.g., CTCA, OCT, IVUS) to enhance segmentation performance by capturing complementary features that may not be visible in a single modality. Additionally, domain adaptation techniques help mitigate performance degradation when applying models across different datasets, scanners, or imaging conditions by aligning feature distributions through adversarial learning, style transfer, or self-training \cite{Qayyum2024_self_training}. Beyond these, unsupervised learning has gained interest as it eliminates the need for labelled data, relying on self-supervised techniques, clustering, or generative models to extract meaningful feature representations \cite{Mazher2024_self_supervised}.

Nonetheless, automated segmentation methods face persisting challenges: 
(1) Image noise, motion blurring from patients' and cardiac motions, and artefacts \textemdash significant coronary calcification and stent implants cause high blooming artefacts in CTCA images, preventing even the latest neural networks from predicting true luminal boundaries \cite{Aljabri2022,Litjens2019}. Similarly, soft plaques have Hounsfield unit values close to the surrounding heart tissues, making it hard for neural networks to differentiate and characterise stenosis severity \cite{Aljabri2022,Litjens2019}. 
(2) Variable image quality \textemdash neural networks usually have the highest predictive efficacy on native images produced by the same scanner following the same protocol as the images used to train them. When applied to external datasets, biased predictions may occur due to over- or underfitting, causing the overall performance to drop \cite{Aljabri2022}. While domain adaptation techniques improve generalisability, their effectiveness remains dataset-dependent and requires further refinement.
(3) Vessel complexity \textemdash coronary arteries are intricate and tortuous, with significant variations and possible congenital irregularities across patients. Algorithms trained upon data without such samples would likely misidentify important coronary features \cite{Gharleghi2022_segmentation_review,Litjens2019}. Few-shot learning has shown promise in addressing this limitation, but its effectiveness depends on the availability of diverse reference samples \cite{huang2025fewshot,liu2024selfsupervised_fewshot}. Therefore, conventional, manual, or at least semi-manual methods are still considered more accurate when the segmentation and reconstruction have been verified by an expert multiple times (intra-observer) or by two or more experts independently (inter-observer). 

Fully automated approaches are expected to accelerate in the coming years, benefiting from this growing openness to publishing and sharing annotated datasets \cite{Gharleghi2023ASOCADATA, li2025medshapenet,Zeng2023_opendata}. Such advances further underpin related machine learning efforts to enhance image quality and artefact handling, including noise reduction and super-resolution methods \cite{Litjens2019}. Therefore, image features of a larger, more diverse population, with variations in scanner vendors and image quality, will assist the required training efforts for future algorithms with improved overall predictive accuracy and generalisability. 

 \FloatBarrier

\section{Computational Analysis of Coronary Haemodynamics}

After data acquisition and image segmentation, CFD simulation can be performed to assess coronary blood flow Figure \ref{CFDsetup}, providing a complete set of haemodynamic parameters further to coronary anatomy and plaque characteristics that are readily measurable from CTCA images. 

A good example of blood flow metrics with diagnostic value is the Fractional Flow Reserve (FFR), which is being widely used in clinical practice. FFR assesses the functional severity of coronary stenoses and thus guides the strategy of Percutaneous Coronary Intervention (PCI). Traditionally, measurement of FFR requires an invasive procedure in which a pressure wire is used to probe the pressure difference across a stenosis under pharmacologically induced hyperaemia. Recent advances in computational modelling have enabled FFR to be calculated from images\cite{FFRct_trial_Norgaard2014, FFRct_trial_Norgaard2015,Kakizaki2022_FFR_OCT,Scoccia2022_FFR_ICA,Yu2021_FFR_IVUS,Takahashi2022_FFR_imaging_comparison}, i.e. $\text{FFR}_{\text{CT}}$, which has proven diagnostic accuracy in various clinical trials  \cite{FFRct_trial_Norgaard2014, FFRct_trial_Norgaard2015}. However, while the performance of $\text{FFR}_{\text{CT}}$ is strong for severe stenoses, its accuracy decreases in moderate lesions \cite{Mittal2023_FFRCT_accuracy,Cook2017_FFRCT_review} and in patients with extensive coronary calcification \cite{mickley2022_FFR_CT_calcification}, making it important to interpret the borderline results with caution. As such, $\text{FFR}_{\text{CT}}$ remains an ideal parameter to be benchmarked against when assessing new computational approaches. Obtaining clinically important haemodynamic metrics through non-invasive CTCA imaging and computational simulations brings genuine advantages to patients,  such as a lower or negligible procedure risk and markedly reduced cost \cite{Douglas2016PLATFORM}. To do this, one should carefully consider appropriate dimensional fidelity, numerical approaches and techniques, boundary conditions, and uncertainty analysis. 

\begin{landscape}
\begin{figure}
    \centering
    \includegraphics[width=1\linewidth]{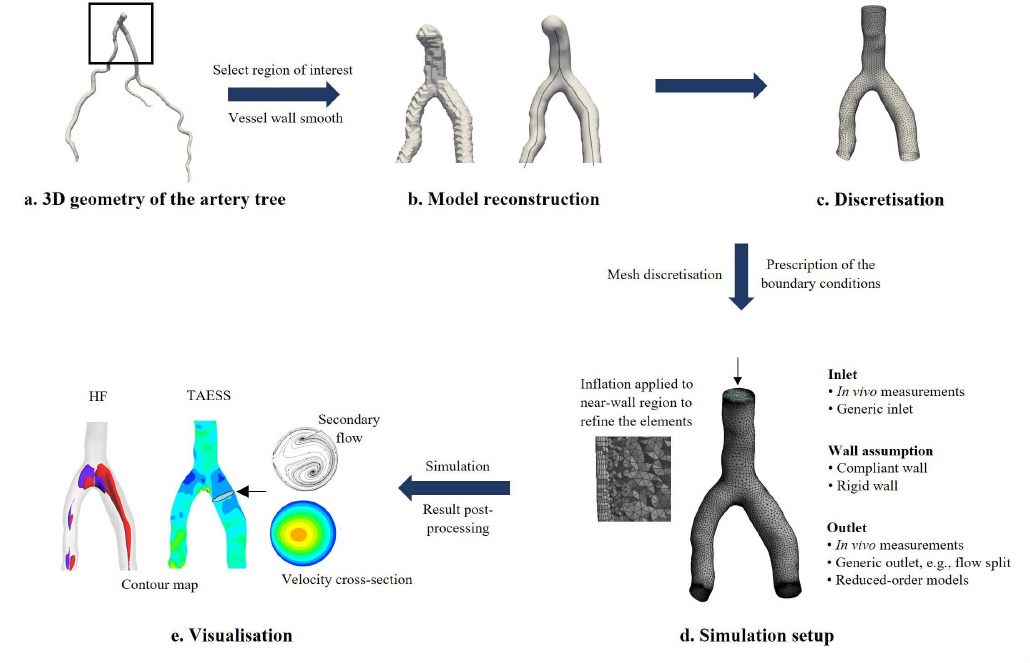}
    \caption{Workflow of a typical CFD analysis for coronary bifurcations. (a) Generation of a virtual representation, often from medical images. (b) Selection of the region of interest and vascular surface smoothing. (c) The smoothed model can be converted to geometry into a computer-aided design file for further modification, such as virtual stent implants. This can be exported as a .STereoLithography (STL) file to be directly imported into CFD software. (d) Computational mesh generation and local grid refinement for the near-wall region. (e) Solving and visualisation of coronary haemodynamics descriptors.}
    \label{CFDsetup}
\end{figure}

\FloatBarrier
\end{landscape}

\subsection{Numerical Approaches}
\begin{landscape}
\begin{figure}
    \centering
    \includegraphics[width=1\linewidth]{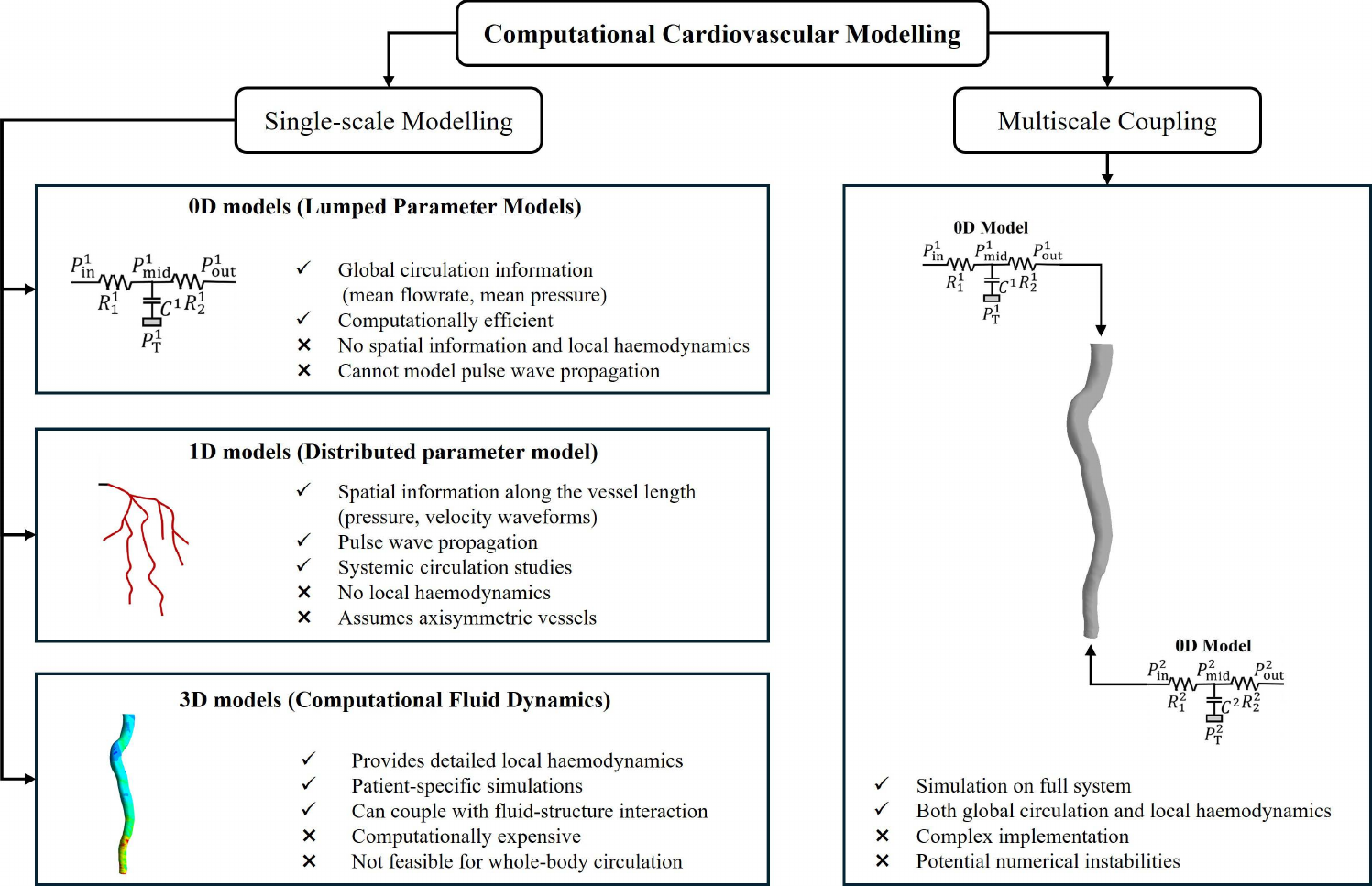}
    \caption{Overview of computational cardiovascular modelling approaches. Single-scale modelling consists of 0D, 1D, and 3D models, each with different levels of spatial dimensions. Multiscale coupling integrates these models to enable comprehensive haemodynamic simulations, balancing computational efficiency with full cardiovascular system consideration. The figures for 0D, 1D, and multiscale coupling were modified from literature \cite{multiscale_fan2024review}, under the terms of the Creative Commons CC BY licence.}
    \label{multiscale}
\end{figure}
\end{landscape}

Coronary blood flow can be modelled with different types of fidelity (Figure \ref{multiscale}), using a zero-dimensional (0D) lumped parameter model, one-dimensional (1D) distributed parameter model, 2D planar model, or 3D full domain model. Lumped parameter models are reduced-order vascular models efficient in quantifying changes in blood pressure, flow, and resistance for a certain length of the blood vessels \cite{Shi2011dimensionality_scales_review}. While it has the lowest computational cost, a lumped parameter model cannot resolve haemodynamic local tensors. 1D blood flow models assume uniform profile, circular cross-section and uniform material behaviour, capable of modelling blood flow with pulse wave transmission along the vessel \cite{Chi2022lumped_parameter_model,Mynard2008ArterialBloodFlowModel,Duanmu2019CoronaryTreeModel}. 2D models characterise the local flow patterns in a plane, representing the 3D flow domain, e.g. when the fluid domains are axially symmetrical. Full-scale 3D simulation accounts for variations in all three spatial dimensions, providing the most detailed blood flow information but requiring the greatest computation cost. Different scales of models can be used independently or combined (Figure \ref{multiscale}). For example, multi-scale models of coronary circulation previously incorporated scale-reduced models (e.g. 0D or 1D models) to then supply dynamic boundary conditions to the full-scale 3D coronary simulations \cite{Lee2012Multi_Scale_Modelling,Keramati2023GaussianProcess,GrandeGutierrez2022HybridModel,Kumar2023PulsatileHemodynamics}. 

Various numerical approaches can be applied to solve the governing equations of blood flow, including the Finite Volume Method (FVM), Finite Element Method (FEM), Lattice Boltzmann Method (LBM), and Smoothed Particle Hydrodynamics (SPH). Open-source packages such as OpenFOAM \cite{openfoam} and SimVascular \cite{simvascular} and commercial ANSYS CFX, Fluent \cite{ansys}, and COMSOL \cite{comsol}, are commonly used. Whilst less popular, LBM prevails in large-scale parallelisation, and SPH is suitable for problems with complex geometries and moving boundaries but has only been applied to LV flow to simulated damaged vessels where its advantages are more appropriate \cite{Toma2021SPH,Topalovic2022SPH,Pontrelli2014LBM}. Comparatively, LBM and SPH require significant expertise to set up using packages like OpenLB \cite{Kummerlander2023_OpenLB} and DualSPHysics \cite{Crespo2015DualSPHysics}. To support clinical and research applications, integrated software solutions with user-friendly platforms also exist, such as the Cardiovascular Angiography Analysis System (CAAS), to provide quantitative analysis of both imaging data and blood flow \cite{Kageyama2023WSSAgreement_CAAS,Tufaro2025FastWSS_CAAS}.

Coronary haemodynamic simulation can be performed in a steady-state manner, quantifying haemodynamics at a specific time point, or in a transient manner, characterising the variations of flow over a cardiac cycle. Depending on the study objective, simulation of the FFR requires only a steady-state fluid dynamic simulation under the hyperaemic condition. However, quantification of time-dependent haemodynamic metrics, e.g. Time-Averaged Endothelial Shear Stress (TAESS), Oscillatory Shear Index (OSI), would require the entire cardiac cycle to be simulated for a couple of periods to minimise the transient start-up effect, and is thus more computationally expensive.

The recent trend in improving computational efficiency is adopting convolutional neural networks to predict the flow parameters of interest, thus allowing to partially or even entirely circumvent solving the governing equations at each time step for a complete fluid flow field \cite{Gharleghi2020ML_WSS_ISBI,Gharleghi2022ML_WSS_journal,AbucideArmas2021,obiols2020ML_CFD,Du2022ML_CFD,Alamir2024ML_WSS}. An emerging neural network model is Physics-Informed Neural Networks (PINNs), which embed governing equations, such as the Navier-Stokes equations, directly into the neural network training process, ensuring solutions that are more consistent with underlying physics \cite{Raissi2019PINN,Cai2021PINNReview}. By integrating physics theory with multi-fidelity and multi-modal data, PINNs demonstrate reduced dependence on large training datasets \cite{Stiasny2023PINN}. PINNs have recently been used to reconstruct velocity fields in aorta, left ventricles and cerebral aneurysms \cite{aghaee2024performance, aghaee2025pinning, shone2023deep, moser2023modeling}  and estimate parameters for reduce-order models \citep{garay2024physics}. Recent studies highlight that while PINNs may be adequate for estimating reduced-ordered models, accurately reconstructing velocity field can be challenging. For example, Aghaee et al. demonstrated that while PINNs could reconstruct velocity field under laminar flow conditions, the accuracy reduced substantially under disturbed and high-frequency flow conditions that can be typical of severe coronary stenosis or aortic coarctation \citep{aghaee2025pinning}. Another recent study by Garay et al. demonstrated that PINNs could be used to reconstruct reduced-order model parameters that may have implications for assigning boundary conditions in coronary CFD simulations; however, similar to the study of Aghaee et al., these authors noted that global velocities could not be reconstructed accurately, especially the vortical structures. Other authors have noted similar observations that highlight challenges of PINNs in reconstructing complex flow patterns \citep{moser2023modeling}. Recent developments in PINNs, such as adaptive sampling, domain decomposition, residual adaptive sampling of PINNs should be explored to improve PINNs accuracy.

Convolutional Neural Networks (CNNs) have been applied to predict the distribution of Endothelial Shear Stress (ESS), also known as Wall Shear Stress (WSS), around coronary bifurcations based on geometric features, achieving high accuracy with deviations of less than 5\% from CFD-derived values while significantly reducing simulation time from 2 hours to under 2 minutes \cite{Gharleghi2020ML_WSS_ISBI,Gharleghi2022ML_WSS_journal}. 

Nonetheless, the accuracy of neural networks relies heavily on the quantity and quality of the training data. Consequently, they inherit uncertainties from prior simulations and cannot generate new metrics beyond what has been simulated. Moreover, rapid estimation of haemodynamic parameters for complex vascular geometries, e.g. beyond the coronary bifurcations, remains challenging. It requires larger-scale training datasets and further network development, before effective application to the entire coronary trees, stented, or stenosed arteries \cite{Li2021ML_CFD,Su2020ML_CFD, Tesche2020MLinCoronaryFlow,Arzani2021BloodFlowPINN}.

\subsection{Numerical Assumptions}

\subsubsection{Rheology}
To solve the governing equations of the fluid flow, i.e. the Navier-Stokes and the continuity equations, the viscosity of the blood should be known. Whilst blood is a non-Newtonian fluid often modelled using the Carreau–Yasuda, Power Law and the Generalised Power Law are deemed necessary for low shear environments to resolve near-wall flow conditions \cite{Soulis2008Non_Newtonian}. However, less computationally intensive Newtonian assumptions may suffice depending on the research question \cite{Abbasian2020blood_model, Johnston2006Non_Newtonian}, and some studies argue that the assumptions on blood rheology have negligible impact on haemodynamics \cite{DeNisco2023rheology}. 

\subsubsection{Wall compliance}
Although the coronary vascular wall is compliant and deformable, it has been modelled as a rigid surface with no-slip condition and without a thickness by the majority of existing studies \cite{BeierBifurcation2016,Kashyap2020curvature,Malve2015Tortuosity}. 
A series of comparative studies support the rigid wall assumption based on agreement between the cycle-averaged near-wall haemodynamics for rigid and deformable vessel walls for both healthy \cite{Torii2009FSI_healthy}, diseased  \cite{Violeta2022FSI_diseased}, or stented coronary arteries \cite{Chiastra2014FSI_stent}. Still, instantaneous haemodynamics was reported to vary between Fluid-Structure Interaction (FSI) and pure CFD \cite{Malve2012FSI_vs_CFD}, especially for the effects on OSI from coronary cyclic movement \cite{Carpenter2023WSSmutidirectionality}, suggesting the necessity of FSI for quantifying the wall deformation and displacement at specific time points in the cardiac cycle. 

Thus, the FSI approach should be used to consider the mutual effect of the vessel and the blood flow and simulate the forces between the fluid and solid domains \cite{Hirschhorn2020FSI_review}. The coupling between the fluid and structural domains can be `unidirectional' (one-way FSI), or `bidirectional' (two-way FSI), whereby for the former, the fluid simulation results are fed to the structural analysis as boundary conditions, but with the latter, the data is either exchanged at each time step (i.e. partitioned FSI), or resolved simultaneously (i.e. monolithic FSI) \cite{Ahamed2017FSI_setup, Hou2012FSI_setup}. 

FSI is more commonly used in stent designs where wall deformation may have a larger effect \cite{Putra2018FSI_Stent,Simao2017FSI_stent,Liu2018stent}, and vulnerable plaque studies \cite{Chhai2017FSI_plaque,Tang2017FSI_plaque}. A key limitation in the field is the lack of understanding of material models to perform high-fidelity FSI. Whilst \textit{in vivo} characterisation of cardiovascular tissues properties is possible through IVUS measurement \cite{Wang2023tissue_property,MasoTalou2018mechanical_characterization}, extensive efforts have been made in an \textit{ex vivo} manner through uniaxial or biaxial loading and indentation tests. Following these tests, parameters in a constitutive model, e.g. a hyperelastic Mooney-Rivilin Model \cite{Cardoso2014mechanical_response}, can be derived to characterise the mechanical response of arterial tissues. However, the number of parameters measured varied across studies and the inter-specimen variation is vast due to inconsistent excision protocols (e.g. post mortem excision time) and experimental conditions (e.g. temperature) \cite{Carpenter2020}. Experimental data indicate that arterial tissue properties are highly nonlinear and anisotropic, to account for which Holzapfel's strain-energy model was proposed to characterise the anisotropic behaviour and collagen fibre orientation, where arterial layers are treated as different fibre-reinforced materials symmetrically disposed with respect to the cylinder axis \cite{Holzapfel2000arterywallmechanics}. Therefore, caution should be exercised when interpreting coronary FSI outcomes. Whilst direct comparison of endothelial shear stress on the vessel wall is not feasible, simulated flow parameters, such as flow velocity and waveform, can be used to indirectly verify the results. For example, in FSI simulations of the aorta, flow parameters were compared against \textit{in vivo} \cite{FSI_in_vivo_Nair2023} and \textit{in vitro} measurements \cite{FSI_in_vitro_Zimmermann2021} to validate the FSI settings. To quantify impact from the variations in material properties, a parametric study on FSI or \textit{in vitro} experiments can be considered, assuming different material properties covering the normal range. Ideally, the resulting blood flow field should further be validated against \textit{in vivo} measurements.

Although limitations exist in coronary FSI, this method is promising for an improved understanding of plaque vulnerability \cite{wu2019FSI_material_properties} and rupture risk\cite{Costopoulos2017plaque_structure_stress}, as FSI is the only simulation approach capable of modelling blood flow under the influence from both the cardiac motion and vascular wall compliance, ensuring a more realistic characterisation of \textit{in vivo} plaque stress and strain \cite{Tang2009MRI_plaque_mechanics}. 

\subsubsection{Inlet and Outlet Boundary Conditions}
Boundary conditions are paramount in coronary CFD analysis, as they directly determine the simulation fidelity. Table \ref{boundary} summarises common boundary conditions for coronary bifurcations simulation. The inlet or outlet conditions can be determined following velocity or pressure data measured \textit{in vivo}. Specifically, \textit{in vivo} velocity data can be obtained by Thrombolysis in Myocardial Infarction (TIMI) frame count \cite{Gibson1996TIMI}, Intravascular Doppler ultrasound \cite{Tanedo2001Doppler}, or continuous thermodilution \cite{Candreva2021thermodilution}. \textit{In vivo} pressure waveforms can be obtained using a pressure wire during an ICA \cite{Mynard2020invivo_pressure_measurements}. However, collecting these data is often challenging and not routinely performed in clinical practice, thus prohibiting large-scale patient-specific studies. 

When \textit{in vivo} data is unavailable, generic boundary conditions can be applied. A common strategy is to scale the waveform or the cycle-averaged inlet flow according to the inlet diameter average at the inlet \cite{ Giessen2011_boundary_conditions,Huo2016scalinglaw,Huo2009VascularVolume}. A flow-split strategy can be applied at each bifurcation to derive the flow rate of the outlet \cite{Giessen2011_boundary_conditions,Huo2009BloodFlowResistance}. The commonly used scaling law and flow split ratio, referred to as Doriot's fit,is based on Murray's Law. This empirical adjustment has led to a robust correlation between flow and diameter at the inlet ($R^2 = 0.87$) and the flow split ratio between daughter branches ($R^2 = 0.70$) when compared to \textit{in vivo} measurements \cite{Giessen2011_boundary_conditions}. Thus, the scaling law and flow split strategy based on Doriot's fit can effectively serve as boundary conditions when \textit{in vivo} data is lacking. One should note that the inlet or one outlet should be set as a free outlet to avoid overconstraining the simulations when using velocity boundary conditions.

Alternatively, reference pressure is commonly applied to the literature outlets. However, pressure or flow rate conditions might fall short in capturing flow redistribution caused by significant stenoses under the hyperaemic condition \cite{zhang2025_flowsplit_reliability}. In such a scenario, multi-scale models with a lumped parameter network characterising the distal microvascular resistance may dynamically produce local haemodynamics close to that measured \textit{in vivo} \cite{zhang2025_flowsplit_reliability}. A well-tuned Windkessel model can produce physiological pressure waveform and values and, thus, fits the \textit{in vivo} measurements better than generic boundary conditions \cite{Pirola2017outlet, Kim2010mutiscale}. 

When using reduced-order models (e.g. lumped parameter models) to provide boundary conditions for 3D coronary simulations, parameter tuning becomes a critical step to ensure physiological reality of the boundary conditions generated \cite{InverseProblemsReview_Nolte_2022,InverseProblemsReview_Nolte_2024}. The tuning process involves solving an inverse problem of identifying a reliable set of resistances and compliances to reproduce flow or pressure waveforms measured \textit{in vivo}. Multiple optimisation approaches exist to refine parameter estimation, including least-squares optimisation \cite{LeastSquares_Cousins_2014,LeastSquares_Carson_2021}, Bayesian estimation \cite{Bayesian_Arnold_2017,Bayesian_Schiavazzi_2017}, and data assimilation techniques \cite{assimilation_DeVault_2008, assimilation_Lal_2017}. Although technical developments such as the recent use of machine learning have automated this process \cite{li2019_lumped_parameter_method}, parameter tuning may still result in local optima, particularly when multiple coupling interfaces exist which significantly expand the parameter space. Incorporating certain flow parameters measured \textit{in vivo} can provide additional constraints to guide parameter estimation \cite{InverseProblemsReview_Nolte_2022}. In the absence of \textit{in vivo} data, flow-split strategies can be used to estimate outlet flow rates to constrain the model \cite{zhang2025_flowsplit_reliability}. 

\begin{table}[ht]
\centering
\caption{Summary of commonly used boundary conditions for coronary computational studies.}
\label{boundary}
\begin{tabular}{| m{1.9cm} | m{2.4cm} | m{7cm} | m{3.5cm} |}
\hline
\rowcolor{headercolor}
\textbf{Boundary} & \textbf{Type} & \textbf{Conditions} & \textbf{Consideration} \\ \hline

\rowcolor{lighteraltrowcolor}
\multirow{2}{4cm}{{Inflow}} & Patient-specific & 
\begin{itemize}[leftmargin=0.3cm]
    \item Velocity or pressure measured \textit{in vivo} (e.g. pressure catheter \cite{Mynard2020invivo_pressure_measurements}, intravascular Doppler ultrasound \cite{Tanedo2001Doppler}, or continuous thermodilution \cite{Candreva2021thermodilution}, TIMI frame count \cite{Gibson1996TIMI})
\end{itemize} 
& Physiologically real but not routinely collected in clinical practice \\ \hhline{~-|-|-}

\rowcolor{lighteraltrowcolor}
& Generic & 
\begin{itemize}[leftmargin=0.3cm]
    \item Flowrates calculated by the vascular diameter \cite{Giessen2011_boundary_conditions,Huo2016scalinglaw,Huo2009VascularVolume}
    \item Static pressure or pressure waveforms from literature \cite{Liu2015_boundary_Influence}
    \item Instantaneous flowrate or pressure values generated by a reduced-order circulatory model (e.g. 0D, 1D) \cite{Pirola2017outlet, Kim2010mutiscale}
\end{itemize} 
& Agrees generally well with \textit{in vivo} measurement \cite{Giessen2011_boundary_conditions} and is easily implemented in large-scale studies \\ \hline

\rowcolor{altrowcolor}
\multirow{2}{4cm}{{Outflow}} & Patient-specific & 
\begin{itemize}[leftmargin=0.3cm]
    \item Velocity or pressure measured \textit{in vivo}
\end{itemize} 
& Physiologically real but not routinely collected in clinical practice \\ \hhline{~-|-|-}

\rowcolor{altrowcolor}
& Generic & 
\begin{itemize}[leftmargin=0.3cm]
    \item Reference pressures (e.g. a 0 Pa pressure at extended outlets when flowrate condition prescribed at inlets) \cite{Liu2015_boundary_Influence}
    \item Flow split at bifurcations (e.g. per the daughter branches’ diameters with an exponent of 2.27) \cite{Giessen2011_boundary_conditions}
    \item Instantaneous flowrate or pressure values generated by a reduced-order (e.g. 0D, 1D) circulatory model \cite{Pirola2017outlet, Kim2010mutiscale}
\end{itemize} 
& Flow split strategy agrees well with \textit{in vivo} measurement \cite{Giessen2011_boundary_conditions}. The reduced-order models capture phasic changes in coronary flow and pressure \cite{Pirola2017outlet, Kim2010mutiscale} \\ \hline

\rowcolor{lighteraltrowcolor}
\multirow{2}{4cm}{{Arterial Wall}} & Patient-specific & 
\begin{itemize}[leftmargin=0.3cm]
    \item Heart-motion-induced arterial bending measured by ICA \cite{Carpenter2021FSI_heart_motion}
\end{itemize} 
& ICA produces only 2D projections. IVUS not routinely performed in clinical practice \\ \hhline{~-|-|-}

\rowcolor{lighteraltrowcolor}
& Generic & 
\begin{itemize}[leftmargin=0.3cm]
    \item Arterial bending assumed by a varying-radius sphere model \cite{Prosi2004curvature_dynamic}
\end{itemize} 
& Arterial bending modelled by the sphere model only verified for the LAD-D1 segments \\ \hline

\end{tabular}
\begin{flushleft}
\footnotesize
Note: For both patient-specific or generic modelling, the arterial wall is always assumed to be rigid, with no-slip condition and no penetration for pure fluid analysis, and its mechanical properties are usually obtained from the literature for fluid-structure interaction analysis.
\end{flushleft}
\end{table}
\newpage

\FloatBarrier

\subsubsection{Verification, Validation, and Uncertainty Quantification}

Computational modelling is a powerful tool for investigating coronary haemodynamics, yet questions regarding its credibility remain \cite{VVUQ_Steinman2018}. Various factors can affect the modelling accuracy, introducing uncertainties across the mathematical and physical modelling process, , including geometry reconstruction, material properties assignment, and boundary conditions specification \cite{VVUQ_Steinman2018,Sankaran2011uncertainty}. To establish the credibility of computational modelling, Verification, Validation, and Uncertainty Quantification (VVUQ) are the recommended processes to follow \cite{Good_Simulation_Practice_Courcelles2024,ASME_guidelines_VVUQ}.

Verification ensures that the computational model correctly implements the underlying mathematical model, assessing numerical accuracy and reducing discretisation errors \cite{Good_Simulation_Practice_Courcelles2024}. One can reduce the uncertainties contributed by the simulation setup by controlling the discretisation procedure. To keep the numerical error in an acceptable range, a fine grid in space and time is necessary \cite{ASME2008discretization}. A mesh convergence test can ensure a fine grid in space \cite{Roache1994mesh,ASME2008discretization}. Computational meshes of complex regions, e.g. stenosis and bifurcation, often need finer meshes to accurately resolve the flow, especially near the boundary, while ensuring numerical stability. For transient simulations, appropriate time step elections and iteration convergence at every time step with discretisation error estimation could reduce the uncertainties \cite{ASME2008discretization}. A mesh dependency test ensures that the spatial and temporal resolutions are fine enough to balance accuracy and computational cost, preventing unnecessary refinements that do not significantly affect results\cite{Sadrehaghighi2021Mesh_Independence_Study}. 

Validation establishes whether the mathematical model accurately represents the reality of interest \cite{Good_Simulation_Practice_Courcelles2024}. One of the main factors influencing validation accuracy is uncertainty in boundary conditions \cite{Ninos2021uncertainty, ASME2008discretization}. In the absence of \textit{in vivo} flow and pressure measurements, one common approach is to use generic boundary conditions scaled to the patient-specific vascular diameter, which might be the only option for assessing the near-wall haemodynamics for large retrospective datasets currently \cite{Candreva2022CFD_review}. However, the generic boundary conditions may not accurately reflect the flow conditions. Even when \textit{in vivo} measurements are available, uncertainties associated with the measurement contribute to the solution's accuracy. For instance, Doppler ultrasound and frame count methods require estimating the cross-sectional lumen area or the hydraulic diameter to calculate the flow rate. The image quality and processing strategies influence the error of the cross-section estimation \cite{Rizzini2022invivomeasurement}.
Moreover, the frame count method is sensitive to image acquisition frame rate \cite{Rizzini2022invivomeasurement}. Although the thermodilution method can directly measure the blood flow rate without the issues of Doppler ultrasound and frame count methods \cite{Rizzini2022invivomeasurement,Gibson1996TIMI}, it requires infusion media, wire intrusion and operation experience, which can potentially introduce new uncertainties \cite{Rizzini2022invivomeasurement,Candreva2021thermodilution}. 

Notably, the boundary conditions, including the generic waveform, lead to a difference in ESS magnitude, which could be decreased after normalisation of ESS \cite{Schrauwen2016boundaryconditions}. Therefore, an improved analysis strategy could reduce the uncertainties of simulations. Standardised simulation setup, boundary conditions and result processing strategy in the future allow for cross-study comparisons before further developing current measurement techniques.

Uncertainty Quantification (UQ) helps identify potential limitations in computational models due to inherent variability or lack of knowledge in input data \cite{Good_Simulation_Practice_Courcelles2024}. Different approaches have been used to quantify the uncertainties in the context of numerical simulation of the arterial blood flow, such as Monte Carlo methods and Gaussian process emulators \cite{melis2017bayesian, tran2017automated, fossan2018uncertainty}. These methods have been used to optimise the boundary conditions \cite{yin2019one} and calculate the sensitivity of a Quantity of Interest (QoI) to the parameters with uncertain or variable values \cite{yin2019one, tran2017automated} to determine the confidence in the calculated. In stochastic methods such as Monte Carlo simulation, the QoI is calculated for numerous random values of the control variables. Therefore, the variation of QoI can be quantified and correlated to the variation of the input parameters \cite{paun2021markov}. The uncertainty quantification using the Gaussian process is through using the information of a limited number of data points and the variation of the interpolation values based on curves derived using the given kernel \cite{yin2019one}. However, since numerous simulations are required to perform such uncertainty quantification analyses, reduced-order models, such as 0D \cite{tran2017automated} or 1D models \cite{yin2019one, paun2021markov}, are more commonly used.

\FloatBarrier

\section{Relevance of Coronary Artery Anatomy and Blood Flow Haemodynamics}

\subsection{Blood Flow, Haemodynamics and Pathophysiological Relevance}

\begin{figure}
    \centering
    \includegraphics[width=1\linewidth]{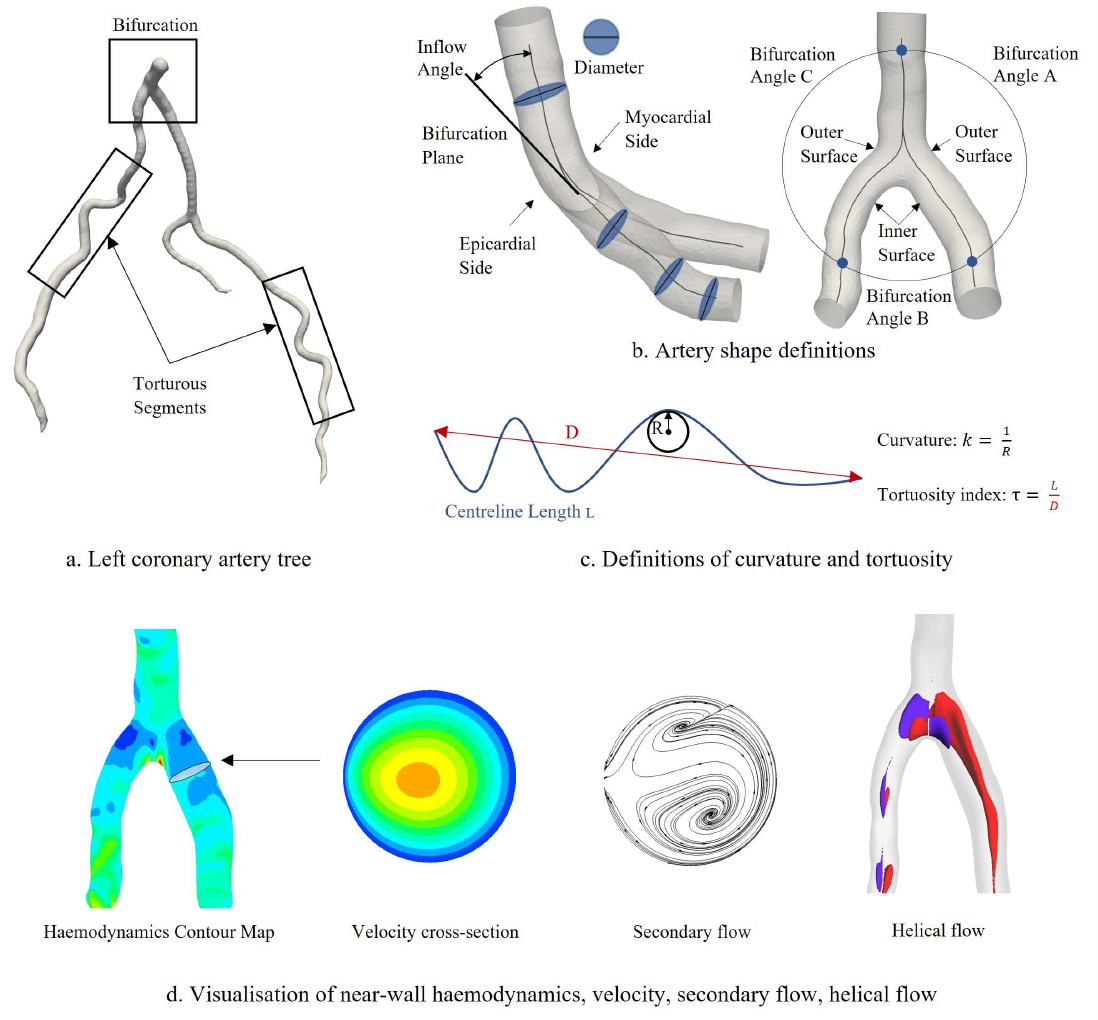}
    \caption{Illustration of shape factors and haemodynamics.(a) A geometry example of a left coronary artery tree. (b) The definitions of coronary shape factors. (c) Numerical definitions of curvature and tortuosity. (d) The visualisation of typical flow patterns in coronary arteries and the near-wall haemodynamics. The 3D simulation used in (d) was performed using ANSYS CFX (ANSYS Inc., Canonsburg, PA, USA).}
    \label{fig:shape_flow}
\end{figure}

CAD is caused by adverse remodelling of the vessel wall due to vascular dysfunction, leading to insufficient myocardial blood supply. Endothelial cells regulate the vascular function in response to the blood flow's mechanical shear forces they are directly exposed to  \cite{Kwak2014Endothelial}. In some instances, adverse stress distributions along the vessel wall caused by disturbed flows can lead to endothelial dysfunction and inflammation, initiating the development and progression of CAD \cite{Timmins2017_WSS_effect}. Thus, the disease is more likely to form in regions with complex anatomical features, where blood flow patterns are affected due to calibre variation and flow path deviations (Figure \ref{fig:shape_flow}).

Several blood flow metrics and flow patterns have been identified as relevant for atherosclerotic plaque development in the coronaries (Tables \ref{descriptors} and \ref{helical_flow}). Haemodynamic wall-based descriptors have been developed to quantify relevant mechanical forces related to endothelial functions (Table \ref{descriptors}), where the most well-established haemodynamic metric is ESS, as suggested by Ku et al. \cite{Ku1985WSS}. Recently, to better differentiate ESS from wall stress, referring to the mechanical stress inside the arterial wall, ESS has been commonly referred to as endothelial stress or ESS instead, a notation that we also adopted here. Low and oscillating ESS promotes a pro-inflammatory state \cite{Kwak2014Endothelial} and remodelling of artery walls \cite{Baeyens2016WSS_effect}, while adversely high ESS is associated with the destabilisation and rupture of plaque \cite{Eshtehardi2017_high_WSS}. However, ESS is a scalar measure representing the magnitude (i.e., length) of the shear stress vector acting on the vessel wall, without capturing directional changes or temporal variations. Other important haemodynamic metrics have additionally been developed previously, including TAESS, transverse ESS (transESS), OSI, Relative Residence Time (RRT) and Topological Shear Variation Index (TSVI). TAESS averages the ESS magnitude over one cardiac cycle. transESS quantifies the component of ESS acting perpendicular to the main direction of blood flow, capturing the multidimensional nature and complexity of shear forces within the vascular system. The OSI quantifies the oscillation in blood flow direction in the vessel \cite{Chatzizisis2008_OSI_basic}, while the RRT is another factor indicating areas with low shear stress and high particle residence time \cite{Hashemi2021_RRT_basic}. High OSI is associated with lipid accumulation and plaque erosion \cite{Luo2021PlaqueErosion,Adriaenssens2021PlaqueProgression}, and high RRT is related to atherosclerotic plaque calcification and necrosis \cite{Kok2019multidirectional_plaqueprogression,Candreva2022CFD_review}. TSVI quantifies the variability of contraction and expansion action of endothelial shear forces along the cardiac cycle \cite{Mazzi2021_TSVI}, with a high TSVI indicating regions of potential future myocardial infarction sites \cite{Candreva2022CFD_review}. In coronary studies, ESS is quantified using its mean value across a specific lumen surface or as its time-averaged value over the cardiac cycle (TAESS). Additionally, the spatial distribution of ESS, OSI, and RRT, is often described as the proportion of the lumen surface exposed to these adverse conditions. This normalisation approach allows for cross-study comparisons across different arterial geometries and patient populations.

\FloatBarrier

\begin{landscape}
\renewcommand{\arraystretch}{1.3}
\begin{longtable}{|p{2.6cm}|m{6.5cm}|m{3.3cm}|m{3.3cm}|m{4.5cm}|}
\caption{The definitions, mathematical equations, threshold values, and clinical relevance of near-wall haemodynamic descriptors related to coronary disease development.}
\label{descriptors} \\

\hline
\rowcolor{headercolor}
\textbf{Haemodynamics Descriptor} & \textbf{Equations} & \textbf{Definitions} & \textbf{Threshold} & \textbf{Clinical Relevance} \\
\hline
\endfirsthead

\multicolumn{5}{c}{{\textit{Continued from previous page}}} \\
\hline
\rowcolor{headercolor}
\textbf{Haemodynamics Descriptor} & \textbf{Equations} & \textbf{Definitions} & \textbf{Threshold} & \textbf{Pathophysiological Relevance and Applications} \\
\hline
\endhead

\hline
\multicolumn{5}{r}{{\textit{Continued on next page}}} \\
\endfoot

\hline
\endlastfoot

\rowcolor{lighteraltrowcolor}
\textbf{Endothelial Shear Stress (ESS)} & $n \cdot \mathbf{\tau}_\text{ij}$ & Shear stress along the vessel wall &  
\makecell[{{m{3.5cm}}}]{Adversely low ESS threshold: \\0.4 Pa \cite{Song2021_tortuosity_stenosis,Chiastra2017Healthy_and_diseased,Kashyap2020curvature,Kashyap2022_tortuosity_definition}, 0.5 Pa\\ \cite{BeierBifurcation2016,Shen2021Secondary_flow,Rabbi2020anatomical_variations}, 0.6 Pa \cite{Malve2015Tortuosity},1.0 Pa \cite{Stone2018PROSPECT}, 1.2 Pa \cite{Koskinas2013LowESS} and tertile-division \cite{Nisco2020HF,Nisco2019HF, Hartman2021NIRS_plaque} \\ Adversely high ESS threshold: 4.71 Pa \cite{Kumar2018highwss,Candreva2022highwss}} & 
\makecell[{{p{4.5cm}}}]{• Normal physiological range: 1.2–1.5 Pa\\ • Low and oscillatory ESS promotes a pro-inflammatory state \cite{Baeyens2016WSS_effect}\\ • Adversely high ESS is associated with endothelial damage, plaque destabilisation, and rupture \cite{Kumar2018highwss, Candreva2022highwss}} \\
\hline

\rowcolor{altrowcolor}
\textbf{ESS Gradient (ESSG)} & $\frac{1}{T} \int_0^T |\nabla \mathbf{\tau}| \, dt$  & Time-averaged ESS gradient & - & Positive ESSG can affect plaque stability \cite{ESSG_Dolan2013} \\
\hline

\rowcolor{lighteraltrowcolor}
\textbf{Transverse ESS (transESS)} & $\frac{1}{T} \int_{0}^{T} \left| \mathbf{ESS} \cdot \left( \mathbf{n} \times \frac{\int_{0}^{T} \mathbf{ESS} \, dt}{\left| \int_{0}^{T} \mathbf{ESS} \, dt \right|} \right) \right| \, dt$  & Description of multi-directionality of ESS during the cardiac cycle & High transESS (tertile-division) indicating large changes in the flow direction and affecting the plaque composition \cite{Kok2019multidirectional_plaqueprogression} & \makecell[{{p{4.5cm}}}]{• Associated with initial plaque growth \cite{Multidirectional_Hoogendoorn2020}} \\
\hline

\rowcolor{altrowcolor}
\textbf{Time-averaged ESS (TAESS)} & $\frac{1}{T} \int_0^T |\mathbf{ESS}|  \, dt$ & The cardiac cycle-averaged ESS & Same as ESS & \makecell[{{p{4.5cm}}}]{• Captures cumulative haemodynamic effects on the vascular wall\\ • More representative of long-term vascular remodelling \cite{Eshtehardi2017_high_WSS}} \\
\hline

\rowcolor{lighteraltrowcolor}
\textbf{Oscillatory Shear Index (OSI)} & $0.5 \left[ 1 - \left( \frac{\left| \int_{0}^{T} \mathbf{ESS} \, dt \right|}{\int_{0}^{T} \left| \mathbf{ESS} \right| \, dt} \right) \right]$   & Representation of the change of the ESS vector from a predominant flow direction & The adverse OSI (> 0.2 \cite{Chiastra2017Healthy_and_diseased} or > 0.1 \cite{zhang2024_new_understanding}) relates to lipid accumulation and plaque erosion \cite{Luo2021PlaqueErosion,Adriaenssens2021PlaqueProgression} & \makecell[{{p{4.5cm}}}]{• Associated with lipid accumulation and plaque erosion \cite{Luo2021PlaqueErosion,Adriaenssens2021PlaqueProgression}} \\
\hline

\rowcolor{altrowcolor}
\textbf{Relative Residence Time (RRT)} & $\frac{1}{(1 - 2 \cdot \mathbf{OSI}) \cdot \mathbf{TAESS}}$  & The residence time of elements in the blood adjacent to the wall & High RRT (> 4.17 $\text{Pa}^{-1}$) is related to atherosclerotic plaque calcification and necrosis \cite{Kok2019multidirectional_plaqueprogression,Chhai2017FSI_plaque} & \makecell[{{p{4.5cm}}}]{• Related to atherosclerotic plaque calcification and necrosis \cite{Kok2019multidirectional_plaqueprogression,Candreva2022CFD_review}} \\
\hline

\rowcolor{lighteraltrowcolor}
\textbf{Topological Shear Variation Index (TSVI)} & $\left\{ \frac{1}{T} \int_{0}^{T} \left[ \nabla \cdot (\mathbf{ESS}_{\text{u}}) - \overline{\nabla \cdot (\mathbf{ESS}_{\text{u}})} \right]^2 dt \right\}^{1/2}$ & The time-averaged root mean square deviation of the divergence of the normalised ESS & Adverse threshold > 40.5 $\text{m}^{-1}$. Higher risk of myocardial infarction \cite{Chhai2017FSI_plaque}  & \makecell[{{p{4.5cm}}}]{• Represents the variability of endothelial shear forces during the cardiac cycle \cite{Mazzi2021_TSVI}\\ • High TSVI indicates potential sites of future myocardial infarction \cite{Mazzi2021_TSVI,Candreva2022CFD_review,Candreva2022highwss}} \\
\hline

\multicolumn{5}{l}{Shear rate: \(\mathbf{\tau}\), local velocity: \( \mathbf{v} \), cardiac cycle: \( T \)} \\

\end{longtable}
\end{landscape}

\begin{figure}
    \centering
    \includegraphics[width=1\linewidth]{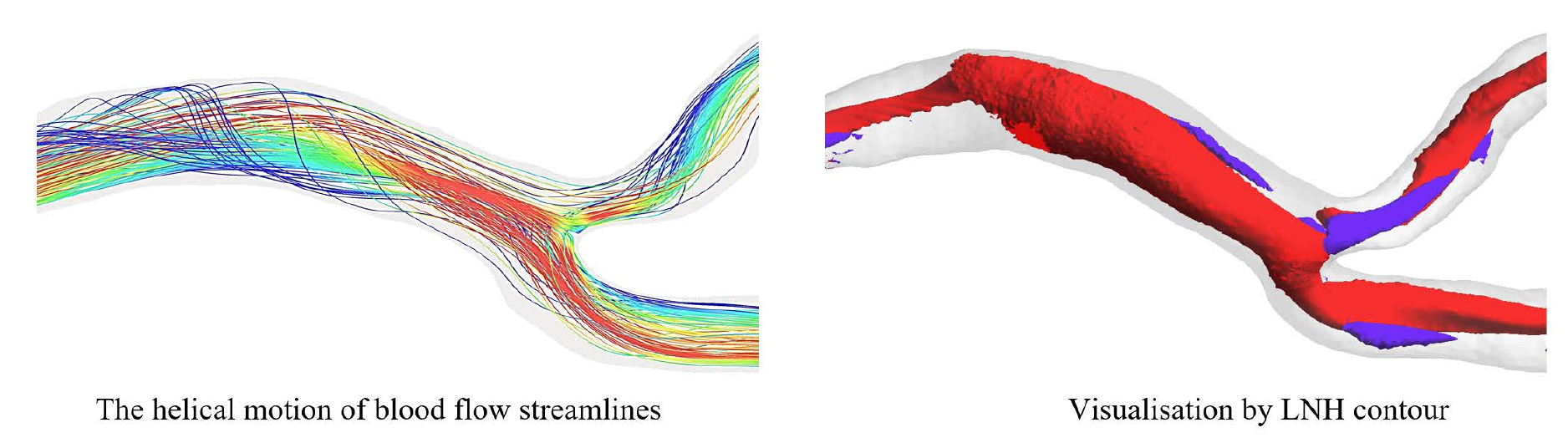}
    \caption{The streamlines for the helical motion of the flow (left) and the corresponding LNH contour (right). The 3D simulations were performed using ANSYS CFX (ANSYS Inc., Canonsburg, PA, USA).}
    \label{fig: helical_fig}
\end{figure}

Similarly, global, non-arterial wall-based metrics include helical, recirculation, secondary, and jet flows. Arterial blood flow under physiological conditions exhibits a helical pattern \cite{Sabbah1984_flow_patterns}. In coronary arteries, the helical flow patterns are characterised by the spiralling of the blood as it moves through the vessels and is affected by the artery shapes. The helical flow can regulate the shear stress distribution on the lumen surface and, thus, has a protective effect against disease formation \cite{Nisco2020HF,Nisco2019HF}. Multiple helicity metrics (Table \ref{helical_flow}) are used to quantify the intensity and rotation features that create helical flow (Figure \ref{fig: helical_fig}). Recirculation happens when the main flow stream encounters obstacles, leading to backward flow and vortices. Secondary flow, initially investigated by Dean in curved pipes \cite{Dean1959}, revealed the secondary flow motion from the outer to inner surfaces in curved segments. The secondary flow pattern in coronary is marked by a pair of counter-rotating vortices \cite{Shen2021Secondary_flow}, which occurs when the flow is split in bifurcations or when it enters curved segments \cite{Ku1997geometry_haemodynamics, Zuin2023secondaryflow}. Currently, there is no established method for quantifying secondary flow in coronary arteries. In computational studies of aneurysm rupture, vortex stability is often evaluated using the Degree of Vortex Volume Overlap (DVO). Given the presence of vortex structures in coronary bifurcations and curved segments, DVO could be explored in future coronary studies to assess vortex stability and its potential role in CAD \cite{Intra_Aneurismal_vortex_Jiang2023, Intra_Aneurismal_vortex_Sunderland2016,Intra_Aneurismal_vortex_Jiang2023}. Jet flow in coronary arteries occurs due to the rapid, localised blood flow acceleration, often caused by narrowing or irregularity of the artery \cite{Ding2021jetflow,Guleren2013stenosis_flow}.

Blood with relatively high velocity in the main coronary vessel impacts the flow divider in bifurcation regions, resulting in asymmetric velocity profiles in the daughter branches (Figure \ref{fig:shape_flow}) \cite{Shen2021Secondary_flow}. This asymmetry is characterised by higher velocities at the inner side near the flow divider and lower ones on the opposite side. This can lead to flow separation, recirculating flows adjacent to the bifurcation point on both sides and secondary flow in downstream daughter branches. Consequently, the bifurcation regions typically show adverse haemodynamics regions at the outer side of the vessels \cite{Antoniadis2015_bifurcation}. In curved segments, curvatures induce directional flow variations due to cross-sectional pressure gradient changes, with higher pressure at the outer side of the curvature. Such changes shift the central fluid layers outward and near-wall fluid inward, creating secondary flows, resulting in low shear stress on the inner side of curvature and high shear stress on the outer side \cite{Shen2021Secondary_flow}. Once the arteries narrowed, the blood passing through narrowed sections experiences rapid velocity and pressure changes resulting in jet flow zone, recirculation, and secondary flow downstream of the stenosis, which can cause varied shear stress distribution there \cite{Molochnikov2023Flow_patterns,Javadzadegan2013_stenosis_recirculation}. Consequently, the blood flow exhibits energy transformations, particularly in kinetic and rotational energy, which have shown predictive capacity for future myocardial infarction \cite{LodiRizzini2024_BloodFlowEnergy}.

\begin{table}[ht]
\centering
\caption{Definitions and mathematical equations of helical flow descriptors for quantification and visualisation.}
\begin{tabular}{|m{3cm}|m{6cm}|m{6cm}|}
\hline
\rowcolor{headercolor}
\textbf{Descriptors} & \textbf{Equations} & \textbf{Definitions} \\
\hline
\rowcolor{lighteraltrowcolor}
$h_1$ & \(\frac{1}{TV} \int_T \int_V \mathbf{v} \cdot \mathbf{\omega} \, dV dt\) & Average helicity \\
\hline
\rowcolor{altrowcolor}
$h_2$ & \(\frac{1}{TV} \int_T \int_V |\mathbf{v} \cdot \mathbf{\omega}| \, dV dt\) & Average helicity intensity \\
\hline
\rowcolor{lighteraltrowcolor}
$h_3$ & \(\frac{h_1}{h_2}\) & Signed Balance of helical flow, \(-1 \leq h_3 \leq 1\) \\
\hline
\rowcolor{altrowcolor}
$h_4$ & \(\frac{|h_1|}{h_2}\) & Unsigned Balance of helical flow, \(-1 \leq h_3 \leq 1\) \\
\hline
\rowcolor{lighteraltrowcolor}
Localised Normalised Helicity
(LNH) & \(\frac{\mathbf{v} \cdot \mathbf{\omega}}{|\mathbf{v}||\mathbf{\omega}|}\) & Visualisation of the orientation of velocity and vorticity vectors \\
\hline
\multicolumn{3}{l}{Local velocity: \( \mathbf{v} \), vorticity vector: \( \mathbf{\omega} \), cardiac cycle: \( T \), arterial volume: \( V \)} \\

\end{tabular}
\label{helical_flow}
\end{table}

\FloatBarrier

Although adverse haemodynamic descriptors correlate to abnormal endothelial function, their translation into clinical settings has not been routinely accomplished. However, this is the focus of ongoing efforts \cite{adikari2022protocol}. The lack of universally accepted values for the predictive haemodynamic descriptors limits their clinical applications. For example, different values have been proposed as the threshold for ESS to be considered adversely low (0.4 Pa \cite{Song2021_tortuosity_stenosis,Chiastra2017Healthy_and_diseased,Kashyap2020curvature,Kashyap2022_tortuosity_definition}, 0.5 Pa \cite{BeierBifurcation2016,Shen2021Secondary_flow,Rabbi2020anatomical_variations}, 0.6 Pa \cite{Malve2015Tortuosity}, 1.0 Pa \cite{Stone2018PROSPECT}, 1.2 Pa \cite{Koskinas2013LowESS}, and tertile-division \cite{Nisco2020HF,Nisco2019HF, Hartman2021NIRS_plaque}). Therefore, testing the prognostic performance of the well-established haemodynamic identifiers through multi-centre randomized control trials will benefit future studies to justify the appropriate threshold values, topological markers, or coronary anatomical characteristics associated with CAD. Future studies would also benefit from considering blood flow metrics with established predictive efficacy for other vascular diseases, such as Low Shear Area and DVO for cerebral aneurysms \cite{Intra_Aneurismal_vortex_Jiang2023, Intra_Aneurismal_vortex_Sunderland2016,Intra_Aneurismal_vortex_Jiang2023}, which share the same theoretical basis to assess the impact on endothelial pathophysiology from blood flow-induced shear stresses.

\subsection{Effects of Anatomical Characteristics}

Many studies have focused on understanding the effect of anatomical variation in coronaries on blood flow (Table \ref{findings} and more details in Appendix Table \ref{supp_table}). Most works focused on the left main bifurcation \cite{Kashyap2020curvature,Kashyap2022_tortuosity_definition,BeierBifurcation2016,Beier2016dynamicsscaling,Gharleghi2025Anatomy_sex_specific,Shen2021Secondary_flow, Morbiducci2016} since it is most commonly treated by intervention. Moreover, standard imaging modalities offer higher clarity for the segment, and the vessel diameter rapidly reduces after the left main \cite{pau2016atlas}. Other bodies of work considered non-bifurcating geometries \cite{Kim2020stenosis,Xie2013tortuosity}, or selected larger segments beyond the left main \cite{Rabbi2020anatomical_variations,Chaichana2011BA}. Very few studies considered the whole tree \cite{Vorobtsova2016tortuosity_BA,zhang2024_new_understanding}. However, 90\% of vulnerable plaques in patients with acute coronary syndromes commonly formed within 42 mm of the left anterior descending artery (LAD) and 40 mm of the left circumflex (LCx) \cite{Araki2020plaque_location}. Spontaneous Coronary Artery Dissection (SCAD) occurs more frequently in the middle to distal coronary segments \cite{Inoue2021scad}. These observations indicate that more comprehensive simulations of the whole coronary artery trees are needed. 

Moreover, anatomical variations within the population are significant \cite{pau2016atlas}. In fact, there is an increasing awareness of anatomical differences between populations \cite{Skowronski2020population_difference}, and the incidence of coronary disease exhibits regional population differences \cite{Shao2020regionaldifference,Ralapanawa2021_population_difference}. Besides, although clinical treatments remain the primary strategy to decrease disease mortality, the population-based prevention approach is more cost-effective, especially for those with low and middle risks \cite{Ahmadi2022populationstrategies}. All these factors highlight the need for large-scale studies to develop more tailored risk assessments.  

\begin{figure}
    \centering
    \includegraphics[width=1\linewidth]{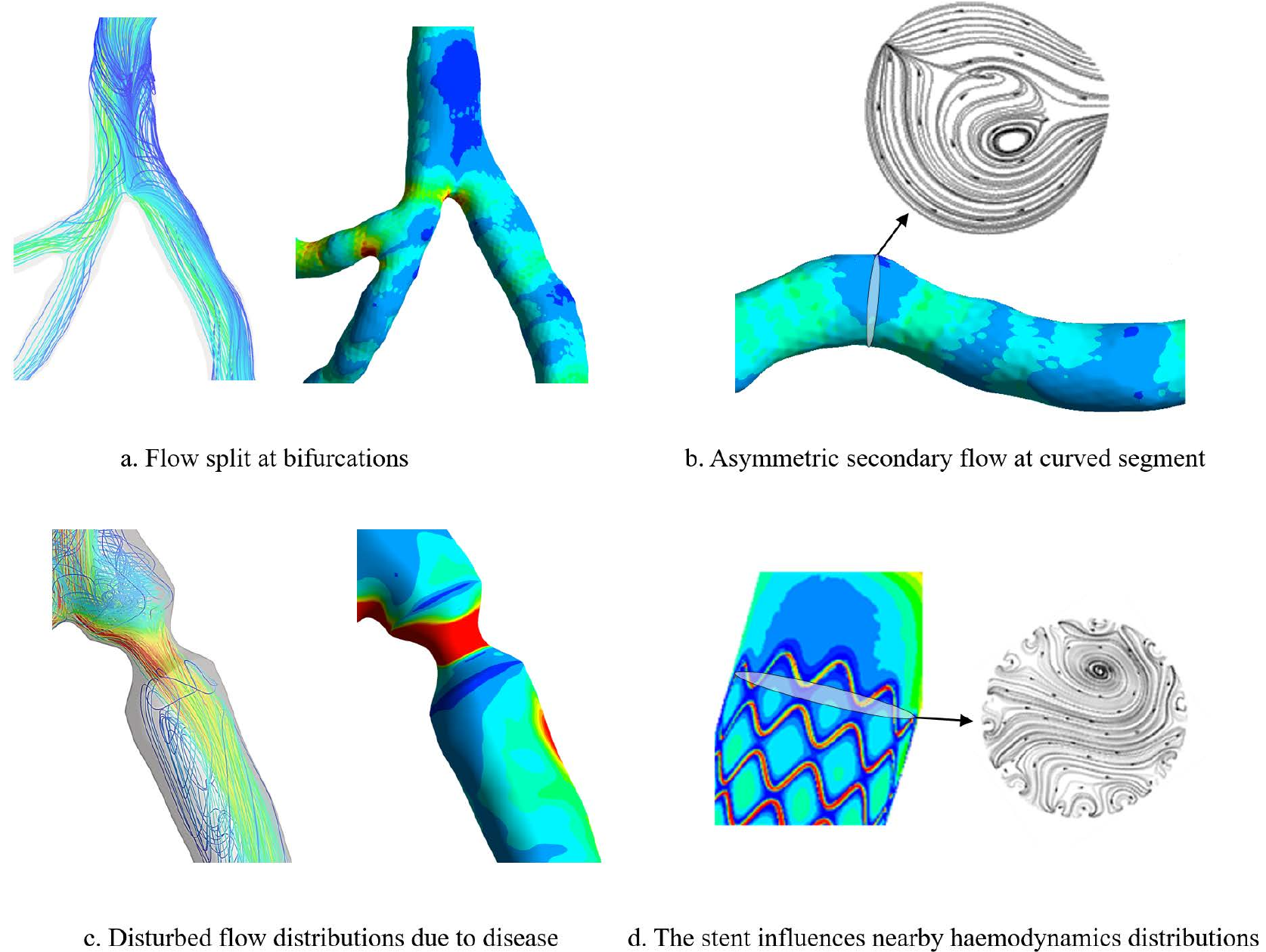}
    \caption{Blood flow streamlines and near-wall haemodynamics distributions influenced by the coronary anatomy and implants. (a) The disturbed flow at bifurcations leads to adverse haemodynamics distributions due to the flow split. (b) The curvature leads to asymmetric secondary flow with coincided low TAESS. (c) The narrowing results in complex flow conditions with low TAESS generated before and after the disease. (d) The existence of stents influences the nearby secondary flow and leads to adverse TAESS adjacent to the struts. The 3D simulations were performed using ANSYS CFX (ANSYS Inc., Canonsburg, PA, USA).}
    \label{last_section}
\end{figure}

Current research is shifting from analysing the impact of a single shape factor to studying multiple factors. Limited investigations studied three \cite{Chiastra2017Healthy_and_diseased} or more anatomical factors \cite{Shen2021Secondary_flow,Gharleghi2025Anatomy_sex_specific,zhang2024_new_understanding}, while most included one or two \cite{Kashyap2020curvature,Vorobtsova2016tortuosity_BA,Song2021_tortuosity_stenosis}. The interdependent effects between the coronary anatomical features \cite{Shen2021Secondary_flow,BeierBifurcation2016} have changed the study direction of the coronary artery to multifactorial analysis. Although the angles of bifurcations were previously regarded as a critical factor  \cite{Chaichana2011BA,Malve2015Tortuosity} due to the distribution of disease around bifurcation regions, the magnitude of the angulations between branches showed a minor influence on haemodynamics compared to the effects of other shape factors \cite{BeierBifurcation2016,Chiastra2017Healthy_and_diseased}. Instead, analyses have shown interdependent effects of bifurcation characteristics, curvature and branch diameter on the haemodynamics distributions \cite{Shen2021Secondary_flow, Huo2012diameter}. Bifurcations alter the path of blood flow through diversion, leading to complex flow patterns such as recirculation and secondary flow, which further generate adverse haemodynamics in nearby regions (Figure \ref{last_section}.a). The large curvature also influences the downstream flow (Figure \ref{last_section}.b), especially with distance to the bifurcation region, causing a skew, spin and asymmetry of secondary flow vortices, increasing helical flow intensity with symmetry loss \cite{Shen2021Secondary_flow,Chiastra2017Healthy_and_diseased, Vorobtsova2016tortuosity_BA}. Adversely high TAESS distributions form after the bifurcations and coincide with the asymmetric flow patterns \cite{Shen2021Secondary_flow}. Besides, diameter and its variation are important shape factors that have yet to be studied in detail. The ratio between the main and daughter branches, which affects the flow split, is defined by Finet's Ratio (FR), i.e. the ratio of the diameter of the inlet over the sum of the diameters of the outlets \cite{Finet2008}. FR strongly correlated with low TAESS ($p<0.001$) \cite{gharleghi2025Anatomy_and_Haemodynamics}. An interdependent effect between diameter and curvature on haemodynamics was reported \cite{Huo2012diameter}. The combined influence of these factors aligns more closely with the distribution of haemodynamics than when considering each factor individually \cite{gharleghi2025Anatomy_and_Haemodynamics}. The interdependent effect of multiple specific anatomical features was also noticed in a study about SCAD, where a combination of the short left main artery, acute bifurcation angle (BA) and strongly curved segments can cause adversely high ESS distribution, which can eventually lead to intimal tear and SCAD \cite{DiDonna2021_Anatomic_features_SCAD}. When considering the entire coronary artery tree, the curvature showed more influences on haemodynamics \cite{zhang2024_new_understanding}. However, further investigations are needed to improve understanding of coronary anatomical feature effects in the artery tree range \cite{zhang2024_new_understanding, Vorobtsova2016tortuosity_BA}. A comprehensive investigation considering the combined effects of various coronary shape factors in the whole artery trees would most robustly determine their correlation with adverse local blood flow. However, considering more shape features complicates the analysis process by increasing the data dimensions, potentially obscuring clear correlations or preventing the extraction of meaningful findings.

Efforts have also been made to examine key anatomical features separately to gain a detailed understanding of their specific impacts. The severely curved vessels have been linked to adverse clinical outcomes, including vessel narrowing \cite{Tuncay2018invivo_tortuosity_CTA}, SCAD \cite{Eleid2014tortuosity} and myocardial ischemia \cite{Li2012_Tortuosity,Vorobtsova2016tortuosity_BA}. However, a contradicting finding showed that 1010 patients with severely curved coronaries showed fewer disease incidences, especially in females 
\cite{Li2011_tortuosity_gender}, which is a potential reason for a lower rate of plaque occurrence in females compared to males \cite{Chiha2017Gender_differences}. In computational studies, tortuous vessels cause larger helicity \cite{Vorobtsova2016tortuosity_BA}, and higher pressure drops \cite{Li2012_Tortuosity}. Some work also found a favourable effect on TAESS due to the high helicity intensity caused by curved segments \cite{Vorobtsova2016tortuosity_BA,Rabbi2020anatomical_variations,Song2021_tortuosity_stenosis}. In contrast, others showed the opposite in idealised and patient-specific non-bifurcating geometries \cite{Xie2014tortuosity}. The contradicting findings may partially be attributed to the sex-specific anatomy and haemodynamics variance \cite{Gharleghi2025Anatomy_sex_specific}. Still, the biggest issue is the measurement matrix, which describes the coronary arteries and is characterised by complex, continuous, curved segments. Previous studies commonly used curvature or tortuosity as a characterisation with inconsistent measurements. Curvature is mathematically defined as the magnitude of the derivative of the unit tangent vector concerning the arc length at a given point. A more intuitive way to define the curvature at a given point is $1/R$, where $R$ is the radius of the osculating circle that approximates the vessel centreline at a point (Figure \ref{fig:shape_flow}.c). The mathematical definition of tortuosity, or tortuosity index, considers the overall path between two points along the curve by comparing the curve length $L$ and the distance between curve endpoints $D$, which cannot capture the spatial information. In clinical practice, the measurements depend on the images and thus are simplified to the number of bendings \cite{Han2022_tortuosity} or their 2D projection as C- or S-shaped \cite{Altintas2020_tortuosity_clinic}. In computational studies, different measurement matrices were used, such as tortuosity index \cite{Malve2015Tortuosity,Vorobtsova2016tortuosity_BA} and average absolute curvature \cite{Malve2015Tortuosity}, or even only considered a single bend using the mathematical definition of curvature, neglecting the spatiality and continuity of this characteristic \cite{Chiastra2017Healthy_and_diseased,Rabbi2020anatomical_variations, Malve2015Tortuosity}. Different measures based respectively on the tortuosity index, absolute curvature, squared curvature, and square-derivative curvature were computed and compared in a recent study, where the mean absolute curvature was recommended due to its strong correlation to the adverse TAESS \cite{Kashyap2022_tortuosity_definition}.

In obstructive diseases (Figure \ref{last_section}.c), the regions narrowed by plaques govern the local haemodynamics \cite{Shen2021Secondary_flow}. The narrowing increases the local flow velocity, causing higher ESS \cite{Peng2016_Geometric_impact_haemodynamics} and higher helicity intensity in diseased regions by an order of magnitude compared to healthy cases \cite{Chiastra2017Healthy_and_diseased}. The increased ESS and TAESS at the site of stenosis most likely to influence plaque composition, leading to destabilisation and increasing the risk of myocardial infarction \cite{Kumar2018highwss, Candreva2022highwss,Eshtehardi2017_high_WSS}. The anatomical feature of the narrowed region has been associated with future myocardial infarction. The Minimum Lumen Ratio (MLR), which describes the abruptness of narrowing in the proximal segment of the lesion, has been identified as a strong indicator \cite{Candreva2024_MLR_MI}. Adverse ESS, TAESS, OSI and RRT form in disturbed flow zones downstream of the narrowing \cite{Pinto2016WSSdescriptors,Frattolin2015_disease,Chaichana2014_disease,Muftuogullari2024_disease,Shen2021Secondary_flow} increasing the risk of accelerating the progression of the obstruction. A similar effect of narrowing was found in hyperaemic conditions \cite{Kamangar2017_disease_hyperemic}. An FSI study showed that the downstream of the stenosis was prone to plaque length progression \cite{Jahromi2019_disease}. Besides, LCx was found to be more atheroprone than LAD when the branch was more than 70\% narrowed \cite{Jahromi2019_disease}. Moreover, the location of stenosis affects the haemodynamics within the whole bifurcation region. Specifically, severe LM-LAD stenosis causes reduced flow in LCx and lower ESS distribution \cite{Peng2016_Geometric_impact_haemodynamics,Frattolin2015_disease,Chaichana2014_disease}. In contrast, LCx stenosis mainly affects the distributions of velocity, ESS, and the pressure gradient of the whole bifurcation \cite{Chaichana2013different_types_of_plaques}.

Stents can also change the haemodynamic quantities (Figure \ref{last_section}.d). Although stents are commonly used to restore blood flow, their structural elements, like rings and struts, can disturb nearby flow. Small and localised recirculation appeared adjacent to the stent struts \cite{Shen2021Secondary_flow,Chiastra2013stent}, which can further influence the existing secondary flow nearby, leading to a more disturbed flow condition \cite{Shen2021Secondary_flow}. Consequently, adverse haemodynamics occur adjacent to the stent struts, potentially developing restensosis in stented regions \cite{Shen2021Secondary_flow}.

\begin{table}[ht]
\centering
\small 

\setlength\extrarowheight{2pt} 
\setlist[itemize]{leftmargin=*} 
\caption{A summary of computed blood flow research relevant to coronary arteries considering flow patterns and shape factors.}

\begin{tabular}{|m{0.2\linewidth}|m{0.74\linewidth}|}
\hline
\rowcolor{headercolor} \textbf{Anatomical factors} & \textbf{Focus and findings} \\
\hline
\rowcolor{lighteraltrowcolor}
\textbf{Multiple of factors} & \begin{itemize}[leftmargin=*]
        \item Complex flow patterns due to the combined effect of the bifurcation structure, continuous curvature and diameter variations  \cite{Shen2021Secondary_flow, Ku1997geometry_haemodynamics, Zuin2023secondaryflow,Antoniadis2015_bifurcation}
    \item Correlation between SCAD and a combination of the short left main artery, small bifurcation angle and strongly curved segments  \cite{DiDonna2021_Anatomic_features_SCAD}
    \item The effect of bifurcation angle is interdependent to other geometrical features and has minimal effects when isolated \cite{BeierBifurcation2016}
  \item Among all investigated features, curvature was a crucial factor for the disease onset  \cite{zhang2024_new_understanding}, strong correlation with the recirculation zone length, \cite{Peng2016_Geometric_impact_haemodynamics,Song2021_tortuosity_stenosis}, while the curvature had a moderate positive \cite{Peng2016_Geometric_impact_haemodynamics} or a negative \cite{Song2021_tortuosity_stenosis} correlation, i.e. having a atheroprotective effect \cite{Rabbi2020anatomical_variations}, curvature effect depends on the local vessel calibre/ diameter variation   \cite{gharleghi2025Anatomy_and_Haemodynamics}
     \item Limited studies considered anatomical features in entire artery trees, while curvature was found to be a crucial  factor impacting disease onset  \cite{zhang2024_new_understanding}

\end{itemize}\\
\hline
\rowcolor{altrowcolor} \textbf{Tortuosity/curvature*} & \begin{itemize}[leftmargin=*]
 \item Unclear effect of the curvature/tortuosity on haemodynamics due to common inappropriate 3D representation \cite{Kashyap2022_tortuosity_definition} and a lack of standardised vessel tortuosity in literature  \cite{Han2022_tortuosity,  Malve2015Tortuosity,Vorobtsova2016tortuosity_BA,  Chiastra2017Healthy_and_diseased,Shen2021Secondary_flow}
  \item Contradictory or inconclusive findings, i.e. large curvature led to high helicity its protective effect \cite{Vorobtsova2016tortuosity_BA}, while linking severe curvature to vessel narrowing \cite{Tuncay2018invivo_tortuosity_CTA}, SCAD \cite{Eleid2014tortuosity} and myocardial ischemia \cite{Li2012_Tortuosity,Vorobtsova2016tortuosity_BA}

\end{itemize} \\
\hline
\rowcolor{lighteraltrowcolor}
\textbf{Narrowing/ Disease} & \begin{itemize}[leftmargin=*]
  \item Stenosis governs local haemodynamics, overriding anatomical effects  \cite{Shen2021Secondary_flow}
  \item High ESS in narrowed regions and adverse haemodynamics (ESS, OSI, and RRT) downstream due to the disturbed flow zones  \cite{Pinto2016WSSdescriptors,Frattolin2015_disease,Chaichana2014_disease,Muftuogullari2024_disease,Shen2021Secondary_flow}
  \item Risk of upstream plaque progression in the proximal area and risk of new plaques formation in the downstream \cite{Jahromi2019_disease}

\end{itemize} \\
\hline
\rowcolor{altrowcolor}
\textbf{Stent} & \begin{itemize}[leftmargin=*]
  \item Stent structure can adversely affect recirculations, the secondary flow and helical flow adjacent to the stent \cite{Shen2021Secondary_flow,Chiastra2013stent}
  \item Additional adverse haemodynamics formed adjacent to the stent struts, potentially leading to further disease development in stented regions \cite{Shen2021Secondary_flow}

\end{itemize} \\
\hline
\end{tabular}
\label{findings}
\begin{flushleft}
\footnotesize
\textbf{*} The definition of tortuosity and curvature overlaps in some literature. Measurements are inconsistent and often inaccurate 3D representation, likely contributing to the contradicting findings. 
\end{flushleft}
\end{table}


\FloatBarrier

\section{Current Limitations and Future Opportunities}

Coronary anatomy is associated with haemodynamics, which influences the formation and progression of CAD. The clinical application of $\text{FFR}_{\text{CT}}$ has indicated that integrating computational methods with imaging techniques can accelerate the translation of haemodynamic insights into more effective prevention and diagnostic strategies. While both $\text{FFR}_{\text{CT}}$ and studies focusing on flow and anatomy utilise haemodynamics to address clinical questions, their applications differ. $\text{FFR}_{\text{CT}}$ is primarily used to assess stenosis severity and guide PCI, whereas studies of coronary anatomy and flow can help to explain the underlying mechanisms of CAD

Despite various \textit{in vivo}, \textit{in vitro}, and \textit{in silico} studies on the association between coronary anatomy, haemodynamics and CAD, the current understanding is still at an early stage and has not yet been translated into routine clinical practice. ESS can now be derived from $\text{FFR}_{\text{CT}}$ analysis using artificial intelligence–enabled quantitative coronary plaque and hemodynamic analysis (AI-QCPHA), which provides a comprehensive lesion-level assessment of high-risk plaque (EMERALD-II trial) \cite{Koo2024AIQCPHA}. However, detailed haemodynamic metrics such as time-averaged measurements or multidirectional haemodynamics such as TSVI still require high-fidelity computational simulations for accurate evaluation. Currently, the computational cost, lengthy segmentation procedures, multiple shape factors from patient-specific geometries, a wide range of shape variations between individuals and the absence of established thresholds for haemodynamic descriptors all limit the understanding of the correlation between coronary artery anatomy and haemodynamics and their effects on CAD development and progression.

Nonetheless, research is increasingly focused on the role of coronary anatomy and blood flow in explaining the higher prevalence of disease in segments with complex geometries \cite{DiDonna2021_Anatomic_features_SCAD,Eleid2014tortuosity,Tuncay2018invivo_tortuosity_CTA}, contributing to assessments of plaque vulnerability \cite{Mazzi2021_TSVI, Multidirectional_Hoogendoorn2020}, informing stenting strategies \cite{Andrea2025_stent_strategies}, and supporting sex-specific risk stratification \cite{Gharleghi2025Anatomy_sex_specific,wentzel2022sex_difference}. This growing body of knowledge presents a promising direction for improving risk prediction, advancing preventative strategies, and enabling more personalised clinical decision-making.

The development of automated segmentation tools has enabled researchers to perform rapid and large-scale segmentation and generate 3D models of the entire coronary artery tree in a feasible time. Therefore, the sample sizes for the previous studies that have used patient-specific models were relatively small. Another limitation is that the segments included in most studies were the main bifurcation regions or non-bifurcating vessels, and most curved segments and some regions commonly susceptible to diseases were neglected. Nevertheless, vulnerable plaques are most commonly located outside the left main region. For example, SCAD occurs in the middle to distal vessel segments. Therefore, investigating the whole tree is essential to understanding the effect of anatomical features on haemodynamics distributions in downstream segments with potential risks \cite{Araki2020plaque_location, Inoue2021scad}. 

Because of all the computational and segmentation challenges, investigations about the relationship between coronary anatomy and haemodynamics with a large population are limited. While patient-specific geometries have increasingly been used, the limited sample size and case selection approach may undermine the generalisability of the analyses, which consequently restricts the clinical applications of current findings. Significant challenges arise from the vast volume of data required and the complexity of analysing multiple shape factors simultaneously to encompass broader patient populations. In this context, power analysis is a valuable tool to determine the appropriate sample size needed to ensure sufficient sensitivity to detect the valid correlations in the whole dataset, thus resulting in efficiency and accuracy in research design \cite{Politi2023sample_size,Hickey2018Politi2023_sample_size}. Future coronary anatomy studies are expected to focus on more extensive and diverse populations, with the representativeness and reliability of their findings supported by their power analysis. Such robust methodology will enhance the validity and applicability of the conclusions and facilitate the integration of the analyses into routine clinical practice.

Due to the shape variations between individuals, there is a clear trend to consider the combined effects of multiple anatomical factors. However, involving multiple factors will increase the data dimension, complicate the analysis, and potentially obscure meaningful correlations. In such scenarios, statistical shape analysis \cite{Ambellan2019_SSM} proposes an effective and practical method to analyse complex shapes and variations, generating representative mean shapes with variations from a group of 3D coronary models \cite{Pau2017shape_in_normal_population}. This approach has previously been used to study heart \cite{Kollar2022_SSM_heart,Suinesiaputra2018_SSM_heart}, aorta \cite{Cosentino2020_SSM_aorta,Wiputra2023_SSM_aorta} or bone shapes \cite{Liu2020_SSM_bone,Ding2022_SSM_bone}. However, although its application in coronary arteries has been tested \cite{Pau2017shape_in_normal_population}, such analyses have valuable potential and are worth further exploration.  

The results and findings of previous studies have proven significant effects of coronary anatomy. However, there are still many limitations that should be addressed. In summary, future studies should consider a few critical points. First, including larger patient-specific coronary datasets with broader population-specific considerations will improve the credibility of the results. Uncertainty quantification is a helpful process that allows researchers to consider the effect of anatomical and physiological variation in their simulations. Second, elucidating the potential interdependent effects of geometrical features on the local values for haemodynamics quantities is necessary for clinically helpful research. Third, developing novel algorithms to automate and accelerate coronary artery segmentation and flow analysis to expedite such studies is essential. 

While this review focuses on coronary arteries, the segmentation and computational pipelines discussed are broadly applicable across the cardiovascular system. Regardless of the vascular regions, three-dimensional geometries can be reconstructed from invasive or non-invasive imaging using the methodologies introduced earlier. By adjusting boundary conditions for different vascular territories, such as the carotid arteries, aorta, and cerebral arteries, CFD modelling can assess near-wall haemodynamics and flow patterns \cite{carotid_Flow_Patterns_elsayed2024assessment, flow_patterns_aorta_shenberger2024effect}, including vortices and recirculation, and their effects on atherosclerosis \cite{carotid_Flow_Patterns_elsayed2024assessment}, aneurysms \cite{Aortic_Aneurysms_manta2025comprehensive,Cerebral_Aneurysms_sherif2024computational,internal_Carotid_Aneurysms_tatsuta2024internal,Extracranial_Carotid_Aneurysms_heeswijk2025extracranial}, dissection \cite{aortic_dissection_chen2013patient}, and post-stent restenosis \cite{stent_colombo2021Femoral,stent_colombo2022Femoral}, as well as the rupture risk of high-risk plaques and aneurysms \cite{aneurysm_rapture_schneiders2015additional}.

Once the link among geometric factors, flow patterns, and haemodynamics descriptors in coronary arteries with large sample sizes is established, clinicians may evaluate the risk of disease development based on vascular physiology and structure during diagnosis, thereby achieving early prevention and reducing the occurrence of severe disease.

\section{Conclusion}
Numerical analyses of the haemodynamics of healthy and diseased coronary arteries have been in the spotlight for years. However, the studies faced numerous limitations due to the anatomical, physiological, and methodological complexities. Concerns about the credibility and reproducibility of computational results still remain. We summarised multiple aspects of these complexities and highlighted the current status of approaches used in relevant investigations. Our review found that larger sample size, interactions between multiple factors, and automation algorithms are the main topics that can address the current limitations. In addition, the standardisation of computational strategies is a primary effort to be pursued to gain credibility and encourage the translation of computational haemodynamics into clinical practice. The budget of uncertainty associated with modelling must be mandatorily delineated, and a comprehensive standardisation strategy should be established. Addressing these challenges will improve the clinical relevance and translation of the studies on coronary artery haemodynamics.

\begin{landscape}
\renewcommand{\thetable}{I}
\section{Appendix}

\begin{longtable}{|p{2cm}|p{4cm}|p{5.5cm}|p{7cm}|p{2.5cm}|}
\caption{Reviewed \textit{in vivo} and computational studies of coronary anatomy and flow. BA: Bifurcation Angle, LMB: Left Main Bifurcation, SCAD: Spontaneous Coronary Artery Dissection, MACE: Major Adverse Cardiovascular Events.} 
\label{supp_table} \\
\hline
\rowcolor{headercolor}
\textbf{Focus} & \textbf{Data} & \textbf{Method} & \textbf{Key Findings} & \textbf{Study} \\
\hline
\endfirsthead

\multicolumn{5}{c}{{\tablename\ \thetable{} -- Continued}} \\
\hline
\rowcolor{headercolor}
\textbf{Focus} & \textbf{Data} & \textbf{Method} & \textbf{Key Findings} & \textbf{Study} \\
\hline
\endhead
\rowcolor{lighteraltrowcolor}
BA, torsion, inflow angle, diameter, curvature, Finet’s ratio & 39 patient-specific left coronary trees based on CTCA & CFD, transient, non-Newtonian Carreau–Yasuda viscosity, scaling law velocity inlet (plug/uniform profile), flow split outlet, rigid wall & - Obstructive coronary disease is correlated with average curvature and normalised area with high OSI. Curvature and OSI may predict plaque onset.  & Zhang et al. (2023) \cite{zhang2024_new_understanding} \\
\hline
\rowcolor{altrowcolor}
BA, inflow angle, curvature, diameter, Finet’s ratio & 127 patient-specific LMB based on CTCA & CFD, transient, non-Newtonian Carreau–Yasuda viscosity model, scaling law velocity inlet (plug/uniform profile), 0 Pa outlet, rigid wall & - Anatomical features (Finet's ratio, diameters and curvatures) are correlated with lowTAESS,  highOSI and highRRT and may be a surrogate of adverse haemodynamics.\newline - Haemodynamic quantities showed high sensitivity to Finet's ration. & Gharleghi et al. (2024) \cite{gharleghi2025Anatomy_and_Haemodynamics} \\
\hline
\rowcolor{lighteraltrowcolor}
Gender, BA, diameter, curvature, Finet’s ratio & 127 patient-specific LMB based on CTCA (12 males and 85 females) & CFD, transient, non-Newtonian Carreau–Yasuda viscosity model, scaling law velocity inlet (plug/uniform profile), 0 Pa outlet, rigid wall & - More area exposed to adverse OSI and RRT in males.\newline - Low TAESS were more present along the inner wall of curvature for females and outer walls for males. & Gharleghi et al. (2023) \cite{Gharleghi2025Anatomy_sex_specific} \\
\hline

\rowcolor{altrowcolor}
Stenosis, tortuosity, curvature, length & 22 left and right coronary arteries based on CTCA & CFD, transient, Newtonian viscosity model, coronary flow rate for inlet, lumped parameter model outlets, rigid, non-slip wall & - The stenosis resulted in formation of recirculation zone at the bifurcation upstream.\newline - High stenosis degree increased the maximum ESS.\newline - The curvature and the length of the lesion segment showed a moderate positive correlation with the length of the recirculation zone. & Peng et al. (2016) \cite{Peng2016_Geometric_impact_haemodynamics} \\
\hline

\rowcolor{lighteraltrowcolor}
BA, curvature, stenosis & 10 idealised LM bifurcations & CFD, transient, non-Newtonian Carreau viscosity model, velocity waveform from literature for the velocity inlet (plug/uniform profile), specified flow split outlet, rigid wall & - BA had a minor impact on haemodynamics in both stenosed and unstenosed models.\newline - Larger curvature moderately increased the area with low TAESS and decreased the area with high OSI and RRT.\newline - Large curvature led to higher helicity intensity.\newline - Helicity intensity in diseased models was one order of magnitude higher than in healthy models. & Chiastra et al. (2017) \cite{Chiastra2017Healthy_and_diseased} \\
\hline

\rowcolor{altrowcolor}
Tortuosity, curvature, stenosis & Idealised non-bifurcating models & CFD, Transient, Newtonian viscosity model, inlet: an eight-coefficient Fourier equation fitted to data from the literature, outlet: constant pressure (13,330 Pa), rigid wall & - Small pulse rate resulted in adverse TAESS. \newline - Severe stenosis increased adverse TAESS distribution. \newline - The tortuosity reduces the area exposed to adverse low TAESS. & Song et al. (2021) \cite{Song2021_tortuosity_stenosis} \\
\hline

\rowcolor{lighteraltrowcolor}
BA, inflow angle, stent & 9 patient-specific and synthetically modified LM models based on CTCA (3 idealised, 3 BA modified, 3 baseline) & CFD, transient, non-Newtonian Carreau–Yasuda viscosity model, scaling law inlet (parabolic profile), 0 Pa pressure outlet, rigid wall & - The BA did not show a significant effect of adverse hemodynamics. \newline  - Presence of stents outweighed anatomical effects. & Beier et al. (2016) \cite{BeierBifurcation2016} \\
\hline

\rowcolor{altrowcolor}
BA, curvature, stent & 128 idealised and patient-specific modified geometries (32 idealised 32 idealised stented 32 patient-specific 32 patient-specific stented) & CFD, transient, non-Newtonian Carreau–Yasuda model, scaling law velocity inlet (parabolic profile), 0 Pa outlet, rigid wall & - Shape factors had an interdependent effect.\newline - Large curvature adversely induces asymmetric secondary flow and low TAESS.\newline - Larger BA reduced areas exposed to low TAESS.\newline -The effect of stents outweighed the effect of BA. & Shen et al. (2021) \cite{Shen2021Secondary_flow} \\
\hline
\rowcolor{lighteraltrowcolor}
BA, diametres & CTCA images from 30 patients (18 males, 12 females, mean age 56 years $\pm$ 8, 22 cases with disease) & Calcified, non-calcified, and mixed plaques, BA between LM and LCx, mean diameter of LAD and Lcx. &- Cases with diseased left coronary arteries had larger mean BA (94° $\pm$ 19.7° VS 75.5° $\pm$ 19.8°, $p = 0.02$). \newline - Cases with diseased left coronary arteries had larger LAD and LCx diameters ($p<0.001$). & Sun and Cao (2011) \cite{Sun2011_tortuosity} \\
\hline
\rowcolor{altrowcolor}
BA, Tortuosity, trifurcation & 20 idealised (10 bifurcations, 10 trifurcations) four patient-specific geometries based on CTCA (3 bifurcations, 1 trifurcation) & CFD, transient, non-Newtonian Carreau viscosity model, pulsatile velocity inlet, constant pressure (80 mmHg) outlets, rigid wall & - Tortuosity induced formation of helical flow, which might be considered an atheroprotective effect due to their effect on near-wall descriptors.\newline - Larger BA and trifurcation angle increased the helicity intensity.\newline - Large branch angles led to more adverse haemodynamics areas, especially in trifurcations. & Rabbi et al. (2020) \cite{Rabbi2020anatomical_variations} \\
\hline

\rowcolor{lighteraltrowcolor}
BA, gender, BMI & CTCA images from 196 patients (129 males, 67 females, mean age 58 $\pm$ 10.5 years) & Calcified, non-calcified and mixed plaques, stenosis degree classification: low (<30\%), intermediate (30-50\%), high (>50\%) risk, BA between LAD and LCx. & - BA>80°was associated with a higher risk of developing CAD. \newline - BMI and gender are significantly correlated with the BA. \newline - Males patients had a 2.07-fold chance of BA>80°.\newline - Patients with high BMI (>25 kg/m$^2$)  had 2.54-fold higher chance of BA>80°. & Temov et al. (2016) \cite{Temov2016bifurcation_CAD} \\
\hline

\rowcolor{altrowcolor}
Tortuosity, curvature, BA, diameter & Eight patient-specific models & CFD, steady, Newtonian viscosity model, scaling law flowrate inlet, flow split outlet, rigid wall & - A significant correlation between the tortuosity index of  LM-LAD segment and low ESS amplitude.\newline - BA was correlated with high ESS amplitude. & Malvè et al. (2015) \cite{Malve2015Tortuosity} \\
\hline

\rowcolor{lighteraltrowcolor}
Tortuosity, bend angle, the length between bends & Non-bifurcating idealised and two realistic LCA models & CFD, transient, Newtonian, waveform from literature for the inlet, constant (0 Pa) pressure outlet, rigid wall & - Adverse ESS existed on the inner wall downstream of the bend with bend angle larger than 120.\newline - Severe tortuosity resulted in abnormal ESS at the bend sections. Therefore, it can be a risk factor for the disease. & Xie et al. (2014) \cite{Xie2014tortuosity} \\
\hline

\rowcolor{altrowcolor}
Curvature, Tortuosity & CTCA images from 73 patients under risk of CAD due to peripheral vascular disease (55 males, 18 females, mean age 63.5 $\pm$ 8.2 years) & Curvature was defined as Menger's curvature, tortuosity was defined as tortuosity index, significant stenosis>70\%. & - Results showed 16.7\% higher curvature for segments with stenosis and 13.8\% higher curvature for arteries with stenosis.\newline - Tortuosity was 30\% higher for segments with plaque, with no significant correlation at the artery level. & Tuncay et al. (2018) \cite{Tuncay2018invivo_tortuosity_CTA} \\
\hline

\rowcolor{lighteraltrowcolor}
Curvature, stenosis & 33 stenoses from 31 patient-specific left coronary arteries based on CTCA & CFD, steady, Newtonian viscosity model, specified velocity based on Re numbers (plug/uniform profile), resistance pressure outlet, rigid wall & - Curvature and the area-reduction and the area-expansion ratios of the stenosis determined the pressure drop. & Kim et al. (2020) \cite{Kim2020stenosis} \\
\hline

\rowcolor{altrowcolor}
Stenosis, BA & Three isealised bifurcations & CFD, transient, non-Newtonian Carreau viscosity, scaling law velocity inlet, 80 mmHg pressure outlet, rigid wall & -Recirculation appeared after carina \newline -High ESS existed in stenosis and carina regions \newline - Post–stenotic LAD and LCx are more likely to have plaque development  & Müftüoğulları et al. \cite{Muftuogullari2024_disease} \\
\hline

\rowcolor{lighteraltrowcolor}
Curvature & 13 Idealised left main bifurcations & CFD, transient, non-Newtonian Carreau-Yasuda viscosity model, velocity waveform inlet (plug/uniform profile),  0 Pa outlet, rigid wall & - Adverse haemodynamics in LAD showed the highest sensitivity to the curvature of other branches.\newline - OSI and RRT showed more sensitivity to curvature compared to TAESS. & Kashyap et al. (2020) \cite{Kashyap2020curvature} \\
\hline

\rowcolor{altrowcolor}
Tortuosity & Three idealised non-bifurcating models, three patient-specific left coronary arteries based on CTCA & CFD, transient, non-Newtonian Carreau viscosity model, scaling law flowrate inlet (plug/uniform profile), constant outlet pressure for idealised cases, lumped parameter resistance model outlet for patient-specific cases, rigid wall & - Tortuosity could have protective effects against atherosclerosis.\newline - Larger tortuosity led to higher helicity intensity.\newline - Helicity is correlated with the pressure drop. An increase in tortuosity may reduce perfusion. & Vorobtsova et al. (2016) \cite{Vorobtsova2016tortuosity_BA} \\
\hline

\rowcolor{lighteraltrowcolor}
Tortuosity & ICA images from 246 patients with confirmed SCAD (45.3±8.9 years old, 96\% female) and 313 patients & \textit{In vivo} \newline  Tortuosity was defined by the numbers of consecutive curvatures (90° to 180°) in segments > 2mm diameter & -High prevalence of tortuosity among patients with SCAD. \newline -Severe tortuosity related to recurrent SCAD & Eleid et al. (2014)  \cite{Eleid2014tortuosity} \\
\hline

\rowcolor{altrowcolor}
Tortuosity & IVUS images from 293 patients & Eccentricity of plaque, lipid, fibrous and calcified plaque, vessel cross-sections were divided into four quadrants.  & - For moderate stenosis, the lipid pool plaque were found to be in the inner curvature and fibrous tissue in the outer curvature.\newline - The results suggests that atherosclerotic lesions progress eccentrically influenced by the blood flow. & Sato et al. (2015) \cite{Sato2015_tortuosity} \\
\hline

\rowcolor{lighteraltrowcolor}
Tortuosity & ICA images from 1010 patients (64±11 year-old, 544 males) & Tortuosity was defined by segments with $\geq$3 bends, and the existence of CAD was defined as stenosis $\geq$50\% & - Severe CAD was located in vessels with low tortuosity ($p = 0.045$).\newline - Large tortuosity was predominantly found in females ($p <0.001$). \newline - Tortuosity showed positive correlation with hypertension and negative correlation with atherosclerosis. \newline - Tortuosity showed no correlation with MACE. & Li et al. (2011) \cite{Li2011_tortuosity_gender} \\
\hline

\rowcolor{altrowcolor}
Tortuosity & ICA images from 870 patients (589 males, 281 females) & Tortuosity was defined as the existence of segments with $\geq$3 bends. Images were analysed by Vessel, Gensini and Extent scores.
 & - Women with severe tortuosity were more likely to have normal coronary arteries than men ($p = 0.03$). & Chiha et al. (2017) \cite{Chiha2017Gender_differences} \\
\hline

\rowcolor{lighteraltrowcolor}
Tortuosity & 127 patient-specific left main bifurcations based on CTCA & CFD, transient, non-Newtonian Carreau–Yasuda viscosity model, scaling law velocity inlet (parabolic profile), constant pressure (0 Pa) outlet, rigid wall & - The average-absolute-curvature had the strongest correlation with the low TAESS and, therefore, was recommended for the measurement of tortuosity. & Kashyap et al. (2022) \cite{Kashyap2022_tortuosity_definition} \\
\hline

\rowcolor{altrowcolor} 
Tortuosity & 21 idealised non-bifurcating models & CFD, steady, Newtonian, 0.156 m/s inlet, 0 Pa outlet, rigid, non-slip wall & - Tortuosity decreased blood pressure \newline - Severe tortuosity may lead to myocardial ischemia. & Li et al. \cite{Li2012_Tortuosity} \\
\hline

\rowcolor{lighteraltrowcolor}
Stenosis & Nine idealised bifurcations with stenosis & CFD, transient, Newtonian viscosity model, inlet velocity waveform scaled by stenosis degree and flow at peak diastole 102.2 mL /min (parabolic profile), 0 Pa outlets, rigid wall & - Existence of stenosis decreased the flow to side branches. \newline - Increased stenosis degree decreased the ESS value. \newline - Stenosis type (1,0,1) showed the greatest risk of ischemia and atherosclerotic growth.
 & Frattolin et al. (2015) \cite{Frattolin2015_disease}\\
\hline

\rowcolor{altrowcolor}
Stenosis & Eight different configurations of plaques in a patient-specific anatomy and case without any plaque  & CFD, transient, Newtonian, velocity inlet from literature, flow split outlets, rigid wall & - Existence of plaque affects the local velocity, pressure gradient and ESS. \newline - Plaque in LCx had the most effects on the velocity, pressure gradients and ESS.  & Chaichana et al. (2013) \cite{Chaichana2013different_types_of_plaques} \\
\hline

\rowcolor{lighteraltrowcolor}
Stenosis & One patient-specific model   & FSI, Newtonian, \textit{in vivo} measurement inlet velocity and outlet pressure waveform, Mooney–Rivlin model for the artery, idealised heart motion & - Areas proximal to the plaque throat had low circumferential strain and low ESS.\newline - The myocardial side was more likely to form disease. \newline - LCx was more atheroprone than LAD when LCx had severe stenosis.   & Jahromi et al. (2019) \cite{Jahromi2019_disease} \\
\hline

\rowcolor{altrowcolor}
Stenosis & One patient-specific LM bifurcation from CTCA, and cases with virtually added occlusion & CFD, transient, non-Newtonian Carreau viscosity model, velocity waveform with specified Re number for inlet, flow split outlet, rigid wall & - Smaller TAESS around bifurcation and along LAD, when Reynolds number was higher. \newline - Low TAESS and high RRT, existed before and after the short stenoses. In contrast, for the long stenoses, such regions occurred on the lateral side of the obstruction.\newline - TAESS value differed between simulations using Newtonian or non-Newtonian blood. & Pinto and Campos (2016) \cite{Pinto2016WSSdescriptors} \\
\hline

\rowcolor{lighteraltrowcolor}
Stenosis & One patient-specific coronary artery (CT) and a diseased case modified from it & CFD, transient, Newtonian, \textit{in vivo} parabolic velocity inlet, flow split outlet, rigid wall & - ESS decreased and wall pressure gradient increased in side branches when plaque existed in LAD and LM. & Chaichana et al (2014). \cite{Chaichana2014_disease} \\
\hline

\rowcolor{altrowcolor}
Stent & Two patient-specific models based on CTCA & CFD, transient, non-Newtonian Carreau viscosity, scaling law velocity inlet (parabolic), flow split outlet, rigid wall & -Region adjacent to the stent struts are prone to restenosis. \newline -Stents resulted in local helical recirculating microstructures. & Chiastra et al. (2013) \cite{Chiastra2013stent} \\
\hline

\rowcolor{lighteraltrowcolor}
BA & CTCA and ICA images from 106 patients with stenosis & CTCA images for anatomical features, ICA for stenosis degree, stenosis degree classification: mild (< 50\%), moderate (50–69\%), severe (70–99\%) and total occlusion (100\%). & -Wider BA was associated with stenosis and non-calcified lesions. & Cui et al. (2017) \cite{Cui2017BA} \\
\hline

\rowcolor{altrowcolor}
BA & Eight idealised bifurcations, four patient-specific left coronary arteries based on CTCA & CFD, transient, Newtonian viscosity model, velocity waveform inlet from literature, pulsatile pressure outlets, rigid wall & - Wider BA led to disturbed flow patterns and low ESSG. & Chaichana et al. (2011) \cite{Chaichana2011BA}\\ 
\hline
\end{longtable}
\end{landscape}

\clearpage
\section{References}

\end{document}